\documentclass[aps,prb,reprint,superscriptaddress,nofootinbib]{revtex4-2}
\usepackage{amsmath}
\usepackage{graphicx}
\usepackage{amsfonts}
\usepackage{color}
\usepackage{bm}
\usepackage{subfigure}
\usepackage{lipsum}
\usepackage{comment}
\usepackage{dsfont}
\usepackage{bm}
\usepackage{amsmath}
\usepackage{amssymb}
\usepackage{graphicx} 
\usepackage{braket}
\usepackage{mathtools}
\usepackage{times}
\usepackage{enumerate}
\usepackage{subcaption}
\usepackage{xcolor}
\usepackage{bbm}
\usepackage[colorlinks,linkcolor=blue,urlcolor=blue,citecolor=blue]{hyperref}

\newcommand{\bea}{\begin{eqnarray}}
\newcommand{\eea}{\end{eqnarray}}

\newcommand{\unit}{1\!\!1}

\begin{document}
 
\title{Coherence-assisted quantum state preparation and thermal control through repeated interactions}

\author{Alessandro Prositto}
\email{alessandro.prositto@mail.utoronto.ca}
\affiliation{Department of Physics and Centre for Quantum Information and Quantum Control, University of Toronto, 60 Saint George St., Toronto, Ontario, M5S 1A7, Canada}

\author{Dvira Segal}
\email{dvira.segal@utoronto.ca}
\affiliation{Department of Chemistry, University of Toronto, 80 Saint George St., Toronto, Ontario, M5S 3H6, Canada}
\affiliation{Department of Physics and Centre for Quantum Information and Quantum Control, University of Toronto, 60 Saint George St., Toronto, Ontario, M5S 1A7, Canada}

\date{\today}

\begin{abstract}
We investigate the relaxation dynamics and steady state of a qubit interacting with a coherent environment through an energy-conserving flip-flop interaction within the Markovian repeated-interaction framework. We derive an exact analytical solution for the evolving and stationary state, nonperturbative in the environmental coherence. We show that environmental coherence provides a resource for dissipative state preparation by expanding the set of accessible steady states. In particular, coherent ancillas can 
increase the system ground-state population beyond that of the bath, corresponding to an effective temperature below the bath temperature,
 and enhance its steady-state purity, with these effects controlled by the interplay between the system energy scale and the system–ancilla coupling. 
Nevertheless, we prove that the purity of the system can never exceed the purity of the ancilla. 
We further analyze the thermodynamics of the protocol and show that environmental coherence can modify and even reverse the natural direction of heat transfer, giving rise to \textit{anomalous heat flow}. Remarkably, for fixed interaction parameters, the direction of the coherence-induced heat transfer is governed by the relative phase between the system and ancilla coherence, whereas their magnitudes determine the amount of heat exchanged. Our results establish environmental coherence as a resource for controlling steady-state preparation and energy transfer in repeated-interaction dynamics.
\end{abstract}

\maketitle


\section{Introduction}
\label{sec:Introduction}

Cooling open quantum systems is a fundamental task in quantum science and technology \cite{Huber2019}. Reaching low temperatures can suppress thermal noise and enhance the performance of quantum sensors \cite{Cappellaro2017,Kim2016}, while the controlled preparation of thermal states of nonzero temperatures is essential for quantum simulations of many-body and open-system phenomena \cite{Abanin2026,Zhu2020}.

Implementing scalable cooling algorithms on digital quantum computers remains a significant challenge, limiting our ability to simulate the finite-temperature dynamics of large, complex systems, including quantum materials and complex molecules \cite{Campisi2024,Polla2021}. This limitation motivates the development of alternative cooling strategies that exploit quantum resources to improve cooling efficiency and scalability \cite{Brunner2014,Laflamme2017, Mahmoudi2020, Scarani2019,Koch2026}.  
In this context, dissipative engineering provides a promising route to state preparation by tailoring the system–environment interaction, the environment itself, or both. Through such engineering, the system can be driven towards a desired target, such as a thermal state, encoded as the steady state of the resulting dynamics \cite{Harrington2022,Cirac2009}. This framework raises two fundamental questions: (i) for a given system–environment setting, which states are reachable through dissipative engineering? (ii) Can the set of reachable states be expanded by introducing additional features into the system–environment interaction or the structure of the environment?
In this work, we address these questions within the framework of repeated-interaction protocols and demonstrate that environmental coherence can serve as a resource for thermal-state preparation. Taking advantage of this coherence as a form of bath engineering, the population of the system can be cooled below the temperature of its environment.

The repeated-interaction (RI) protocol, also known as the quantum collision model, provides a simple and versatile framework for investigating and simulating open quantum systems beyond the regimes typically accessible to conventional quantum master equations, such as the weak-coupling regime \cite{Rau1963,RevRI,Merkli2014}. In its basic Markovian formulation, the environment is represented by a sequence of identical, noninteracting, and uncorrelated ancillas. The system interacts sequentially with each ancilla for a fixed duration $\tau$, with no free evolution between interactions. After each interaction, the ancilla is discarded and replaced by a fresh copy, and the process is repeated.

RI protocols have been extensively used to investigate concepts in quantum thermodynamics \cite{RevRI,Barra2015,Parrondo2022,Paternostro2019,Gabrielle2026} and quantum information processing \cite{Buzek2002PRA,Zhang2025}. More recently, they have emerged as a promising framework for simulating open quantum systems on digital quantum computers \cite{Carollo2023,Pocrnic25}. The sequential structure of RI protocols naturally maps onto gate-based quantum circuits, providing a direct route for implementing open-system dynamics on quantum hardware. In particular, RI models offer an intuitive description of thermalization processes, including cooling and heating, and can inspire algorithmic strategies for thermal-state preparation on digital quantum computers \cite{Hagan2025, Abanin2026, LinLin2026, Prositto2025Dynamics,CRE2025,Prositto2026, Eisert2012}.

Building on early studies of thermalization in RI models with partial-swap interactions \cite{Buzek2002PRA,Buzek2002PRL} and subsequent numerical investigations beyond this regime \cite{Campbell20}, Ref.~\citenum{Prositto2025Dynamics} demonstrated that engineering the system–ancilla interaction can enlarge the set of accessible steady states. While partial-swap interactions drive the system toward thermal equilibrium at bath temperature, introducing asymmetry through an anisotropic XY interaction enables the preparation of thermal states at temperatures different from, but always higher than, the bath temperature. This result was established {\it analytically} through a complete characterization of the relaxation dynamics and steady state of a minimal Markovian qubit–qubit collision model with an anisotropic XY interaction.

However, Ref.~\citenum{Prositto2025Dynamics}, like much of the collision-model literature \cite{RevRI}, focuses on thermal environments, in which ancillas are prepared in Gibbs states that are diagonal in the energy eigenbasis and characterized by a fixed temperature. In contrast, RI dynamics in the presence of coherent environments remains comparatively unexplored: The role of coherence in collision models has previously been addressed in Refs.~\citenum{Landi2019, DeChiara2024,DeChiara2022,MPaternostro2017,Paternostro2019}, yet the analysis was restricted to the perturbative regime of coherence and carried out within the framework of a Lindblad master equation. 

A broad body of work has established quantum coherence as a genuine quantum resource, enabling processes and performance advantages that are inaccessible to incoherent states \cite{Gour2022,Chabaud2025}. 
More broadly, coherence can constitute a form of \textit{athermality}, a resource associated with deviations from thermal equilibrium. These concepts have been formalized within quantum resource theories \cite{DeChiara2022,Spekkens2013,Dana-Winter2017,Winter-Yang2016,Eisert2017,Plenio2014} and have found applications in quantum information and computing \cite{Plenio2026,PRX2021,Ng2024,Plenio2022,Chabaud2025}, quantum sensing \cite{Yadin2025,Ullah2023}, and quantum thermodynamics \cite{Narasimhachar2021,Jennings2016,Huber2015,Aimet2023, Seah21,Haack23}. Coherence has also been identified as an important ingredient in complex phenomena in biological \cite{Lloyd2011,Gregory2019,Chin2012} and chemical systems \cite{Scholes2017,Vitali2022}. Understanding how coherence is generated, transferred, and harnessed is therefore a fundamental problem in quantum dynamics, motivating this study on how environmental coherence modifies RI dynamics and the resulting steady states.

In this work, we investigate the relaxation dynamics and steady state of a single qubit interacting sequentially with a stream of ancilla qubits within a minimal Markovian RI model governed by a flip-flop interaction. In contrast to the conventional thermal setting, the ancillas are prepared in quantum states that include coherence. Exploiting the simplicity of the RI model, we achieve an exact analytical solution for the relaxation dynamics without imposing any restriction on the magnitude of the ancilla coherence (beyond demanding physical density matrices with hermiticity, unit trace, and positive semidefiniteness).
This enables us to explore regimes beyond weak-coherence approximations and to systematically characterize how environmental coherence modifies both the relaxation dynamics and the resulting steady state. 

Our main contributions are as follows.

(i) We derive an exact analytical solution for the dynamics and steady state under environmental coherence. We show that bath coherence acts as a quantum resource, expanding the set of accessible steady states. In particular, it enables cooling below the bath temperature and enhances steady-state purity.

(ii) We show that coherence alone is not sufficient to produce these effects. Its ability to induce cooling and enhance steady-state purity depends crucially on the interplay between the relevant energy scales of the model, set by the system–ancilla coupling strength and the qubit frequency. 

(iii) We show that environmental coherence can control the direction of heat transfer between the system and ancilla, even driving heat against the temperature gradient and thus giving rise to anomalous heat flow \cite{Partovi2008,Lu2026}. The direction of the coherence-induced heat flow is governed by the relative phase between the system and the ancillary coherences.

The paper is organized as follows. In Sec.~\ref{sec:model}, we introduce the model and derive its exact analytical solution.
In Sec.~\ref{sec:cooling_and_purification}, we investigate the cooling below bath temperature and the enhancement of steady-state purity induced by environmental coherence. In Sec.~\ref{sec:thermodynamics}, we analyze the thermodynamics of the model, including the emergence of anomalous heat flow. In Sec.~\ref{sec:conclusions}, we summarize our results and discuss future work.

\section{Model}
\label{sec:model}

We consider a two-level system (qubit) that sequentially interacts with a stream of ancillary qubits (ancillas) via a flip-flop interaction described by the Hamiltonian
\begin{equation}
\label{eq: flipflop_interaction_Hamitlonian}
    \hat{H}_I = J(\hat{\sigma}_x^S\otimes\hat{\sigma}_x^A+\hat{\sigma}_y^S\otimes\hat{\sigma}_y^A).
\end{equation}
Here $\hat \sigma_{\alpha}^{\bullet},\:\alpha=x,y,z$ stands for the Pauli operators in the Hilbert space of the system (S) or ancilla (A).
Throughout this study, we work with units of $\hbar\equiv 1$ and $k_B\equiv1$ for the Boltzmann constant. 

The total Hamiltonian of the model, $\hat{H}$, is given as the sum of the interaction Hamiltonian $\hat H_I$ and the free Hamiltonian
\begin{align}
    \hat H_0 = \hat H_S \otimes \unit_A +  \unit_S\otimes \hat H_A,
    \label{eq:totalfreeHamiltonian}
\end{align}
where
\begin{equation}
    \label{eq: general_system/ancilla_Hamiltonia}
    \hat{H}_{\bullet} = -\frac{\omega}{2} \hat{\sigma}_z^{\bullet},
\end{equation}
with $\omega$ as the frequency of the qubit. 
The symbol $\unit_\bullet$ represents the unit operator in the Hilbert space of the system or the ancilla.
We set the system and the ancilla qubits to be resonant, $\omega_S=\omega_A\equiv\omega$. In this setup, it is easy to check that $[\hat H_I,\hat H_0]=0$, which implies that the dynamics is energy-conserving. 
We assume each ancilla to be prepared in a fixed identical initial state, 
\begin{equation}
    \rho_A = \begin{pmatrix}
        p_A &c_A\\
        c_A^{*}& 1-p_A
    \end{pmatrix},
    \label{eq:ancilla_densitymatrix}
\end{equation}
with $p_A$ the ground state population and $c_A$ non-zero coherence. We also assume that there are no interactions or initial correlations among the ancillas. After interacting with the system, the ancilla is discarded to never interact again, or equivalently, refreshed to its initial state. This implies that the generated RI dynamics is Markovian and that no memory effects between collisions are present.
 
We allow an arbitrary initial state for the system. Formally, at step $n$, the state of the system is described by 
\begin{equation}
    \rho_S^{(n)} = \begin{pmatrix}
        p_S^{(n)}&c_S^{(n)}\\
        (c_S^{(n)})^{*}& 1-p_S^{(n)}
    \end{pmatrix},
\end{equation}
where $p_S^{(n)}$ denotes the ground state population of the system and $c_S^{(n)}$ is the coherence.
The time evolution of the system's state is described by the RI map,
\begin{align}
\label{eq:map}
    \rho_S^{(n+1)} = \text{Tr}_{A}\left[\hat U(\tau) \rho_S^{(n)}\otimes \rho_A \hat U(\tau)^{\dagger}\right],
\end{align}
with the unitary $\hat U(\tau)=e^{-i\hat{H}\tau}$ generated by the total Hamiltonian $\hat H$. 
This model has been analytically solved in the regime of weak coherence in Refs. \citenum{Landi2019,DeChiara2024,DeChiara2022}. Here, we solve the dynamics without any assumptions on the magnitude of the ancillas' coherence.

\subsection{Thermal ancillas}

\begin{figure}[ht!]
    \centering
\includegraphics[width=1\linewidth]{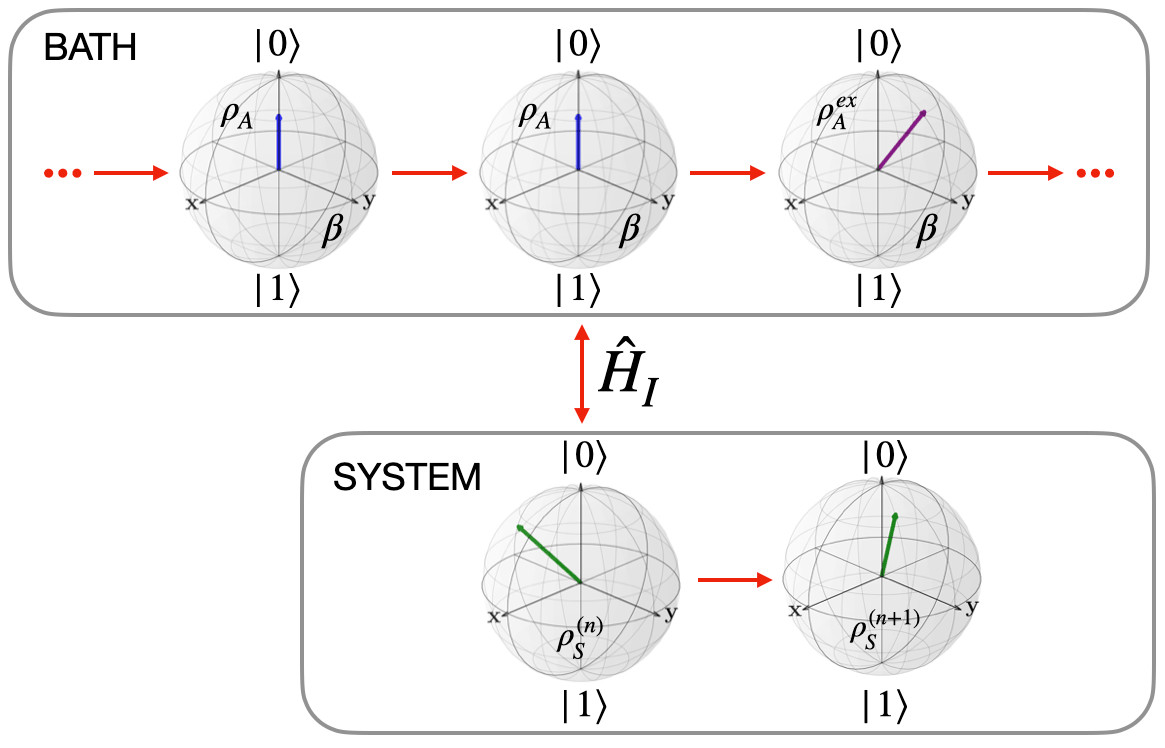}
    \caption{Diagram of the dynamics induced by sequential interactions with thermal ancillas at temperature $1/\beta$. At each step $n$, the system evolves from  state $\rho_S^{(n)}$ to state $\rho_S^{(n+1)}$ through an interaction with an ancilla in the Gibbs state $\rho_A$. Due to the interaction, the ancilla evolves into an ``exhausted" state $\rho_A^{ex}$. The ancilla is replaced or refreshed before the next interaction. The system evolves towards the thermal state of temperature $1/\beta$.
    }
    \label{fig:ThermalBath_Diagram}
\end{figure}

We first review the thermal bath case.
If the ancillas are thermal, $c_A$ = 0, the dynamics can be easily solved, yielding \cite{Prositto2025Dynamics}
\begin{align}
    p_S^{(n+1)} &= p_A+\cos^{2}(2J\tau)\left(p_S^{(n)}-p_A\right),\nonumber\\
    c_S^{(n+1)} &= \cos(2J\tau)e^{i\omega\tau}c_S^{(n)},
\label{eq:thermalancillas_n_step}
\end{align}
leading to the steady state $(n\to \infty)$,
\begin{align}
p_S^{(\infty)} = p_A, \,\,\,\,
c_S^{(\infty)} = 0.   
\label{eq:thermal_ancillas_ss}
\end{align}
The system thus approaches the thermal state of the ancilla in the asymptotic infinite-collision limit.
In Fig. \ref{fig:ThermalBath_Diagram}, we depict a collision of the system qubit with a thermal ancilla, leading to degradation of the coherence of the system.

\subsection{Coherent ancillas}
\label{sec:ModelB}

\begin{figure}[ht!]
    \centering
    \includegraphics[width=1\linewidth]{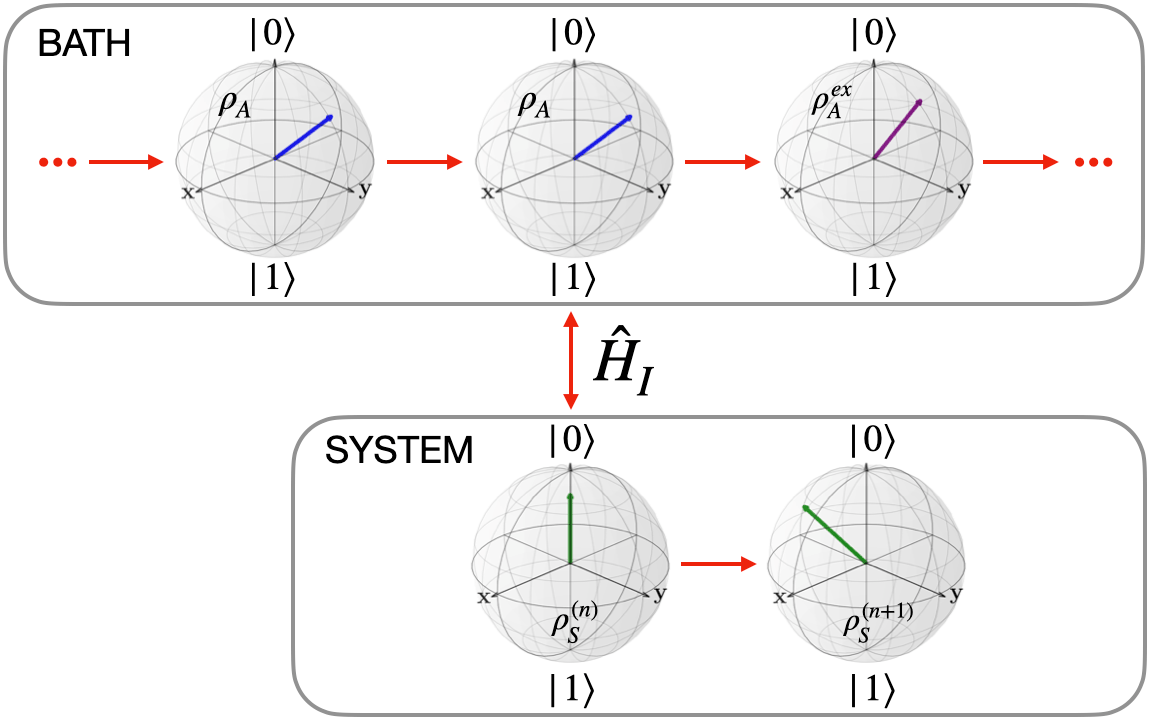}
    \caption{Diagram of system dynamics induced by sequential interactions with coherent ancillas. At each  step $n$, the system evolves from state $\rho_S^{(n)}$ to  $\rho_S^{(n+1)}$ through an interaction with a coherent ancilla in the initial state $\rho_A$. The ancilla  evolves into an ``exhausted" state $\rho_A^{ex}$. It is replaced or refreshed before the next interaction. The system evolves towards its fixed point, a state with nonzero coherence.}
    \label{fig:CoherentBath_Diagram}
\end{figure}

We keep the same Markovian map (\ref{eq:map}), but now allow for coherence in the ancilla's state. 
A schematic representation of the RI model with coherent ancillas is shown in Fig.~\ref{fig:CoherentBath_Diagram}. By solving the dynamics exactly for a general coherent ancilla state, we obtain
\begin{align}
p_S^{(n+1)} &= p_A+ \cos^{2}(2J\tau)\left(p_S^{(n)}-p_A\right)
\nonumber\\&-\sin(4J\tau)\text{Im}\left(c_S^{(n)}c_A^*\right) ,\label{eq:CoherentAncilla_exactsol_pop_nstep}\\
c_S^{(n+1)} &=e^{i\omega\tau}\left[c_S^{(n)}\cos(2J\tau)+i\left(2p_S^{(n)}-1\right)c_A\sin(2J\tau)\right].
\label{eq:CoherentAncilla_exactsol_coh_nstep}
\end{align}
The population and coherence are no longer evolving independently, as they were in the thermal case. We discuss this point in detail in Appendix \ref{app:population_coherence}. 
As for the steady state solution, we get
\begin{align}
p_S^{(\infty)} &= p_A\nonumber\\& +\frac{2|c_A|^2}{K}(2p_A-1)\cos(2J\tau)\Big[\cos(2J\tau)-\cos\left(\omega\tau\right)\Big] \label{eq:coherentancilla_exact_pop_ss},\\
c_S^{(\infty)} &=- i \frac{c_A}{K}(2p_A-1)\sin(2J\tau)\Big[\cos(2 J \tau)-\cos(\omega\tau)\Big]\nonumber\\
&-\frac{c_A}{K}(2p_A-1)\sin(2J\tau)\sin(\omega \tau),
\label{eq:coherentancilla_exact_coh_ss}
\end{align}
with
\begin{align}
    K&= K(J,\tau,\omega,|c_A|)\nonumber\\ &= \big[\cos(2J\tau)-\cos(\omega\tau)\big]^2+\sin^2(\omega\tau)\nonumber
    \\&-4|c_A|^2\cos(2J\tau)\big[\cos(2J\tau)-\cos(\omega\tau)\big].
    \label{eq:coherentancilla_ss_denominator}
\end{align}
In Appendix  \ref{app:cohancillas_eom}, we report the corresponding differential equations of motion derived in the limit $\tau\rightarrow0^+$. Furthermore,  in Appendix \ref{app:K_nonnegative}, we establish that $K$ is always nonnegative, i.e., $K \geq 0$. 
An equality $K=0$ can occur but only at specific parameter values, $\cos(\omega\tau)=\cos(2J\tau)=\pm1$, for which the RI map, in fact, does not evolve,
$p_S^{(n+1}=p_S^{(n)}$ and $c_S^{(n+1)}=c_S^{(n)}$, thus
there is no unique steady state for these values. We exclude these solutions because they do not lead to dynamics, and thus we use $K>0$.

%

The expressions for the transient and steady-state behavior of the system are central to our analysis. Importantly, the solution is {\it nonperturbative} in the ancilla coherence $|c_A|$ and does not rely on any assumption regarding the relative magnitudes of the system-ancilla coupling strength $J$ and the system frequency $\omega$, contrasting common assumptions made in the derivation of the corresponding Lindblad limit \cite{RevRI}. 

In what follows, we define ``cooling'' as the regime in which the steady-state ground-state population of the system exceeds that of the ancillas, $p_S^{(\infty)}>p_A$. As we show below, such cooling can be enabled by coherence in the ancilla states.
 
In steady state, the ground state population could be either higher or lower than the ancilla's, depending on the sign of the function $f(\tau)=\cos(2J\tau)\big[\cos(2J\tau)-\cos(\omega\tau)\big]$, see Eq. (\ref{eq:coherentancilla_exact_pop_ss}), thus on the relative magnitude of the frequency $\omega$ to the coupling strength $J$.

We now simplify the steady state result to provide more insights.
We define
\begin{align}
D &\equiv  \left[\cos(2J\tau)-\cos(\omega\tau)\right]^2
+\sin^2(\omega\tau), \nonumber\\[0.3em]
\Lambda &\equiv
\frac{
4|c_A|^2\cos(2J\tau)
\left[\cos(2J\tau)-\cos(\omega\tau)\right]
}{D}, \label{eq:Dlambdadef}
\end{align}
which yields $K=D(1-\Lambda)$. Note that $D>0$ since we ignore non-relaxing points as discussed below Eq. (\ref{eq:coherentancilla_ss_denominator}).
With these definitions, it can readily be shown that the relative polarization is
\bea
\frac {2p_S^{(\infty)}-1}
{{2p_A-1}}= \frac{1}{1-\Lambda}.
\label{eq:pss}
\eea
Thus, coherence modifies the system's polarization by a multiplicative factor. The case  $\Lambda>0$ leads to enhanced polarization (cooling), while $\Lambda<0$ means reduced polarization (heating). Recall that we work under the assumption of no population inversion for the ancilla, $p_A>1/2$.

In the perturbative limit,
\begin{align}
p_S^{(\infty)}
&\simeq p_A\nonumber\\&
+ \frac{
2|c_A|^2(2p_A-1) \cos(2J\tau)
\left[\cos(2J\tau)-\cos(\omega\tau)\right]
}{ D } 
\nonumber\\
&+O(|c_A|^4).
\end{align}
This shows that the leading correction to the steady-state population is quadratic in the magnitude of the environmental coherence, $|c_A|^2$.

As for the steady-state coherence, we simplify Eq. (\ref{eq:coherentancilla_exact_coh_ss}) using polar coordinates. We define
\bea
\sqrt{D}e^{i\theta} = \sin (\omega\tau) + i(\cos(2J\tau)-\cos(\omega\tau))
\eea
leading to 
\bea
c_S^{(\infty)} = -\frac{c_A(2p_A-1)\sin(2J\tau)}{\sqrt{D}(1-\Lambda)} e^{i\theta}, 
\eea
or equivalently,
\bea 
c_S^{(\infty)} =
-c_A \left(2p_S^{(\infty)}-1\right)
\frac{\sin(2J\tau)}{\sqrt{D}} e^{i\theta}.
\label{eq:css}
\eea
Focusing on the magnitude of coherence,  $|c_S^{(\infty)}|$, this relation connects the  coherence of the system to three quantities: environmental coherence $|c_A|$, the resulting steady-state system polarization $|2p_S^{(\infty)}-1|$, and a factor determined by Hamiltonian parameters and the collision time, $J$, $\omega$, and $\tau$.
In the perturbative regime,
\bea c_S^{(\infty)}
\simeq -c_A(2p_A-1) \frac{\sin(2J\tau)}{\sqrt{D}}
e^{i\theta} +O(|c_A|^3).
\eea
Unlike the system's population, which appears at order
$|c_A|^2$, the induced system coherence is linear in the ancilla coherence at its leading order.

Equations 
(\ref{eq:CoherentAncilla_exactsol_pop_nstep})-(\ref{eq:CoherentAncilla_exactsol_coh_nstep}) for the transient evolution, and Eqs.
(\ref{eq:pss})-(\ref{eq:css}) for the steady state are the main analytical result of our study, providing clean expressions for the study of cooling and purification, as we discuss next.  We highlight that no weak coupling, short collision time, or any assumption on the relative magnitude of $\omega$ and $J$ were made.

\section{Cooling and purification by coherent ancillas}
\label{sec:cooling_and_purification}

A key consequence of the coupling between population and coherence dynamics,
Eqs.~(\ref{eq:CoherentAncilla_exactsol_pop_nstep})-(\ref{eq:CoherentAncilla_exactsol_coh_nstep}),
is the possibility of cooling the system below the bath temperature. For thermal
ancillas, Eq.~(\ref{eq:thermal_ancillas_ss}) shows that a flip-flop system-ancilla interaction yields $p_S^{(\infty)}=p_A$, such that the system
thermalizes to the bath temperature. In Ref.~\citenum{Prositto2025Dynamics},
we showed that introducing anisotropy into the system--ancilla interaction,
for example through an anisotropic Heisenberg interaction (anisotropic XY model), allows the
steady-state temperature of the system to be tuned. However, for thermal ancillas, this tuning remains bounded by $p_S^{(\infty)}\leq p_A$, and therefore
a thermal ancilla cannot cool the system below the bath temperature, even when the interaction requires external work. 

Here, we demonstrate that introducing environmental coherence removes this restriction, allowing cooling of the system {\it below} bath temperature, i.e., $p_S^{(\infty)}>p_A$. 
Throughout, ‘cooling’ refers to an increase in the population of the ground-state, or equivalently, a decrease in the effective temperature defined by the population.
Whether such cooling occurs depends on the interplay between the relevant energy scales of the model, namely, the system-ancilla coupling strength $J$ and the qubit frequency $\omega$. We further show that population cooling necessarily enhances the steady-state purity, whereas enhanced purity can also be achieved in the absence of cooling.

Since our aim is to isolate the role of coherence in cooling, we restrict our analysis to $p_A>1/2$, corresponding to ancillas with a finite positive effective temperature. We therefore exclude population-inverted environments, $p_A<1/2$, for which population inversion may itself constitute a thermodynamic resource \cite{Sparaciari2017,Barra2019}. We also exclude the special case $p_A=1/2$, which corresponds to infinite temperature. At this point, the system relaxes to the maximally-mixed state, with $p_S^{(\infty)}=1/2$ and $c_S^{(\infty)}=0$, thus eliminating any steady-state effects associated with the ancillary coherence.

\begin{figure}
    \centering
    \includegraphics[width=1\linewidth]{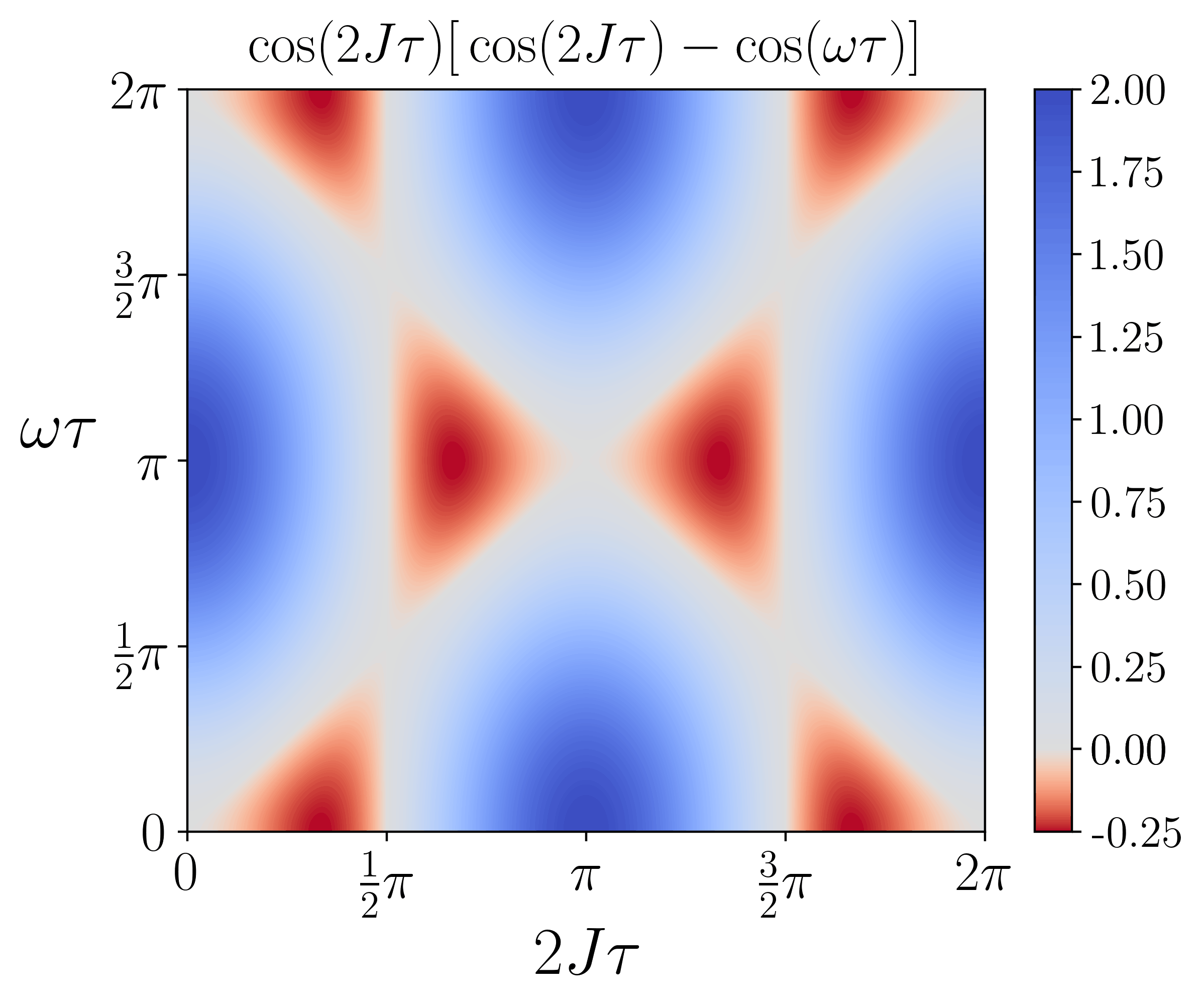}
\caption{The cooling condition, $f(2J\tau,\omega\tau) = \cos(2J\tau)\big[\cos(2J\tau)-\cos(\omega\tau)\big]>0$, see Eq. 
(\ref{eq:coolingcond}).
Regions in blue indicate enhancing the ground state population of the system relative to the ancilla (cooling); regions in red represent domain where the opposite is true (heating).}
\label{fig:cooling_parameterregion}
\end{figure}

\subsection{Cooling by coherence}

Environmental coherence can enable cooling below the bath temperature, in the sense of enhancing the ground state population of the system above the ancilla's, $p_S^{(\infty)}>p_A$. According to
Eq.~(\ref{eq:coherentancilla_exact_pop_ss}), this condition reduces to
\begin{equation}
    \cos(2J\tau)\big[\cos(2J\tau)-\cos(\omega\tau)\big]>0.
    \label{eq:coolingcond}
\end{equation}
Recall that $p_A>1/2$ and $K> 0$. 
In Fig.~\ref{fig:cooling_parameterregion}, we plot 
Eq.~(\ref{eq:coolingcond}), identifying the parameter regions in which cooling occurs. The steady-state behavior is governed by the relative magnitudes of
$J$ and $\omega$. The cooling regions are symmetric about $2J\tau=\pi$. In general, cooling occurs in two distinct regimes:
(i) weak system-ancilla coupling, $2J\tau\ll\pi/2$, combined with a large
spin splitting, $\omega\tau\approx\pi$; and (ii) the complementary regime of
small spin splitting, $\omega\tau\ll\pi/2$, combined with strong system-ancilla coupling, $2J\tau\approx\pi$.

We also identify particular values of $J$ and $\omega$ that produce unique behaviors. For example, when  $\cos(2 J \tau) = \cos(\omega\tau)$ we get $p_S^{(\infty)} = p_A$. Namely, the system achieves the ancilla's population despite the presence of environmental coherence.
However, when $\omega \tau =2(n\pi-J\tau)$ steady-state coherences are not canceled out, $c_S^{(\infty)} = -c_A(2p_A-1)\sin^2(2J\tau)/K$, and thus the system is not thermal.
Another special case occurs when $2J\tau = (2k+1)\pi$
with $k$ an integer.
Then, $c_S^{(\infty)}\equiv0$ and $p_S^{(\infty)}\geq p_A$. This means that at this point, the system cannot be heated above the temperature of the bath and can only be cooled. 

In particular, when $\cos(2J\tau)=-\cos(\omega\tau)=-1$, Eq.~(\ref{eq:coherentancilla_exact_pop_ss}) reduces to 
\begin{align}
p_S^{(\infty)} &= p_A+\frac{|c_A|^2}{1-2|c_A|^2}(2p_A-1)\nonumber\\
&\leq p_A+\frac{p_A(1-p_A)}{1-2p_A(1-p_A)}(2p_A-1),\label{eq:max_popss}
\end{align}
being $|c_A|^2\leq p_A(1-p_A)$ by positivity. To quantify the cooling, we consider the ratio $\left[\frac{p_A(1-p_A)}{1-2p_A(1-p_A)}(2p_A-1)\right]/p_A$. Maximizing this ratio with
respect to $p_A$ provides that the maximum cooling in terms of a higher population of the ground state is reached when $p_A\approx\sqrt{2}/2$, where we find that the system is cooled to approximately $20.9\%$ of the ancilla population. 

Moreover, when $\cos(2J\tau)=-\cos(\omega\tau)=-1$, we get $\Lambda=2|c_A|^2$. It then follows from Eq.~(\ref{eq:coherentancilla_exact_pop_ss}) that the steady-state system polarization, $2p_S^{(\infty)}-1$,
is enhanced by the ancilla coherence, and, for maximal coherence, $|c_A|^2=1/4$, it reaches twice the ancilla polarization.

This highlights the role of environmental coherence as a quantum resource: even in the absence of initial system coherence, coherence in the environment can be consumed to cool the system in the sense of increasing the ground state population.

\subsection{Purification does not imply cooling, but cooling implies purification}

Purification is a central task in quantum science and technology \cite{Bennett1997, WildeBook, NielsenChuangBook}.
It consists of driving a quantum system from a mixed or noisy state toward a pure state. Recent studies have investigated purification in multilevel systems \cite{Vyas2025}, purification through repeated measurements \cite{Dasari2022,Sen2022}, and fundamental physical limits on the achievable degree of purification \cite{Brumer2013,Ticozzi2014}.
Here, we show that environmental coherence can serve as a resource for increasing the steady-state purity reached by a system under RI dynamics, yet without ever exceeding the purity of the environmental ancilla. We demonstrate that purification can accompany cooling below the bath temperature, but can also occur independently of cooling, including  when the system is heated above the bath temperature. Thus, in our model, cooling below the bath temperature is sufficient but not necessary for purification.

\begin{figure*}[ht]
    \centering
    \includegraphics[width=1\linewidth]{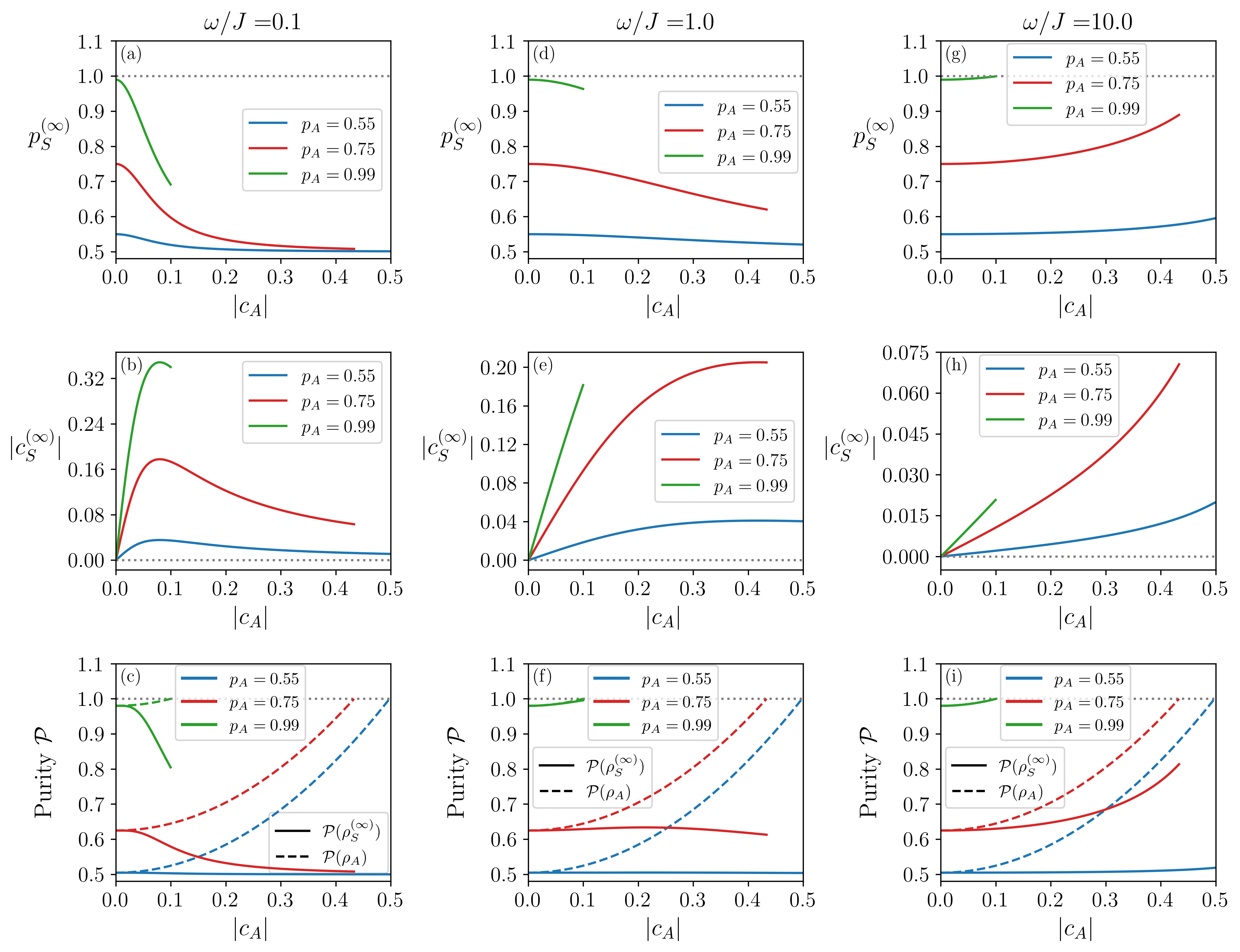}
    \caption{Impact of environmental coherence on populations (a,d,g), coherence (b,e,h), and purity (c,f,i) of the system at steady state, reported for different values of the ratio $\omega/J$. In (c,f,i), we also show, as a dashed line, the purity of the environmental ancilla before a collision. We set $J=1$ and $\tau=0.1$, varying $\omega\in\{0.1,1,10\}$. We selected three values for $p_A$, i.e. the ground state population of the ancilla, varying the coherence magnitude in $\left(0,\sqrt{p_A(1-p_A)}\right)$ to preserve the positivity of the state. We generated this panel using the exact steady-state solution of the model, Eq.~(\ref{eq:coherentancilla_exact_pop_ss}) and Eq.~(\ref{eq:coherentancilla_exact_coh_ss}).}
    \label{fig:CoherentAncillaImpactOnSteadyState}
\end{figure*}

The steady-state purity of a qubit can be formally written as
\begin{align}
    \mathcal{P}\Big(\rho_S^{(\infty)}\Big) &=\text{Tr}\Big[(\rho_S^{{(\infty)}})^2\Big]\nonumber \\
    &=(p_S^{(\infty)})^2+(1-p_S^{(\infty)})^2+ 2|c_S^{(\infty)}|^2.
    \label{eq:purityqubit_formal}
\end{align}
Defining the population difference in the ground state, between the steady state and that of the ancilla, $\Delta p = p_S^{(\infty)}-p_A$, we  write  
\begin{align}
    \mathcal{P}(\rho_S^{(\infty)}) &= (p_A+\Delta p)^2+(1-p_A-\Delta p)^2+ 2|c_S^{(\infty)}|^2\nonumber\\&=p_A^2+(1-p_A)^2\nonumber\\&+2\Delta p^2+2\Delta p (2p_A-1) +2|c_S^{(\infty)}|^2, 
    \label{eq:PrhoS}
\end{align}
where we recognize $p_A^2+(1-p_A)^2$ as  
the steady-state purity reached by the system in the absence of environmental coherence.
Note that this quantity differs from the purity of the coherent ancillas, which is $\mathcal{P}(\rho_A)=p_A^2+(1-p_A)^2+2|c_A|^2$.

Cooling the system below the bath temperature implies $\Delta p >0$.  
We then note that
\begin{align}
    2\Delta p^2+2\Delta p (2p_A-1) +2|c_S^{(\infty)}|^2>0,
    \label{eq:purity_inequality}
\end{align}
since we assume $p_A>1/2$. Inspecting Eq. (\ref{eq:PrhoS}), this implies that the steady-state purity of the system is higher than what would have been obtained with thermal ancillas, for which the system would have relaxed to a diagonal thermal state.
In other words, environmental coherence increases the purity of the system in steady state {\it when there is a cooling process}.

However, an improvement in system purity in steady state can also be observed in the absence of cooling: Eq.~(\ref{eq:purity_inequality}) can be satisfied even when $\Delta p\leq0$. For example, if we set $\cos(2J\tau) =\cos(\omega\tau)$, i.e., $\omega=2J$, we have  $p_S^{(\infty)} =p_A$. At that point, there is no cooling, but there is an increase in the purity of the system compared to that of the thermal case.

It remains to determine whether the steady-state purity of the system can {\it exceed} that of the coherent ancillas. In Appendix \ref{app:proof_DeltaPurity}, we prove analytically that this is never possible: for both thermal and coherent ancillas, 
\bea
\mathcal{P}(\rho_S^{(\infty)})\leq\mathcal{P}(\rho_A). 
\eea
Moreover, equality is achieved only for thermal ancillas, $c_A=0$. In this case, in fact, the steady state of the system coincides with the state of the ancilla and therefore it reaches the same purity.

This result is consistent with the broader literature on the fundamental limits of quantum purification and cooling, which shows that the degree of purification attainable by a system is constrained by the resources available in its environment and by the allowed system-environment dynamics \cite{Ticozzi2014}. However, to the best of our knowledge, explicit studies testing such limits \cite{Koch2021}, including the ones in the context of heat-bath algorithmic cooling \cite{Mosca2015,Laflamme2016,Raeisi2021}, have only considered thermal (Gibbs) ancillas. Here, we extend this picture to ancillas with the state including coherence, showing that coherence can modify the stationary state of the system, yet it cannot drive the system into a purity exceeding that of the ancillary systems that generate the dynamics. 

We report numerical results regarding cooling and purification in Fig. \ref{fig:CoherentAncillaImpactOnSteadyState}, analyzing the steady state and its purity relative to the thermal case for three different choices of $\omega/J$, 0.1, 1 and 10. 

When $\omega\ll J$ (e.g., $\omega=0.1$, $J=1$), we find that the ancilla coherence enhances the transfer of energy into the system: the additional term originating from the environmental coherence in the population of the system, see Eq.~(\ref{eq:coherentancilla_exact_pop_ss}), becomes negative, leading to $p_S^{(\infty)}<p_A$ and hence to heating of the system population {\it beyond} what would be induced by a thermal bath alone (a). 

At the same time, the ancilla coherence is transferred to the system. Indeed, Eq.~(\ref{eq:coherentancilla_exact_coh_ss}) is of the form $|c_S^{(\infty)}|=A|c_A|/(B+C|c_A|^2)$, with $A,B,C\geq0$,
for the parameters of Fig. \ref{fig:CoherentAncillaImpactOnSteadyState} (a)-(c), 
explaining the non-monotonic behavior observed in Fig. \ref{fig:CoherentAncillaImpactOnSteadyState}(b): increasing environmental coherence initially enhances the system coherence, which reaches a maximum before decreasing towards its asymptotic value. Despite this coherence transfer, the system purity decreases even while the ancilla purity increases with $|c_A|$ growing, see Fig. \ref{fig:CoherentAncillaImpactOnSteadyState}(c), showing that in this regime ancilla coherence enhances coherence generation  - and heating - leading to reduced purification overall.

A qualitatively different behavior emerges when $\omega\sim J$ (e.g., $\omega=J=1$). The system is still heated above the temperature of the bath, showing as decreasing ground state population with $|c_A|$, see Fig. \ref{fig:CoherentAncillaImpactOnSteadyState}(d), but the coherent ancillas can now do more than merely generate coherence in the system: for $p_A$ close to 1, they also increase the purity reached by the system; see the green line in Fig. \ref{fig:CoherentAncillaImpactOnSteadyState}(f). 

Strikingly, in the regime $\omega\gg J$ (e.g., $\omega=10$, $J=1$), environmental coherence reverse their effect: the second term in Eq.~(\ref{eq:coherentancilla_exact_pop_ss}) becomes positive, and the system is {\it cooled} below the bath temperature, see Fig. \ref{fig:CoherentAncillaImpactOnSteadyState}(g). Thus, the same environmental coherence that enhances heating at small $\omega/J$ becomes a resource for cooling at large $\omega/J$. In this regime, the system coherence increases monotonically with $|c_A|$, Fig. \ref{fig:CoherentAncillaImpactOnSteadyState}(h), accompanied by an increase in system purity, Fig. \ref{fig:CoherentAncillaImpactOnSteadyState}(i). These results reveal a non-trivial control mechanism: by tuning the ratio $\omega/J$, environmental coherence can be converted from a heating source into a cooling and purification resource.

\begin{figure}[ht!]
    \centering
\includegraphics[width=1\linewidth]{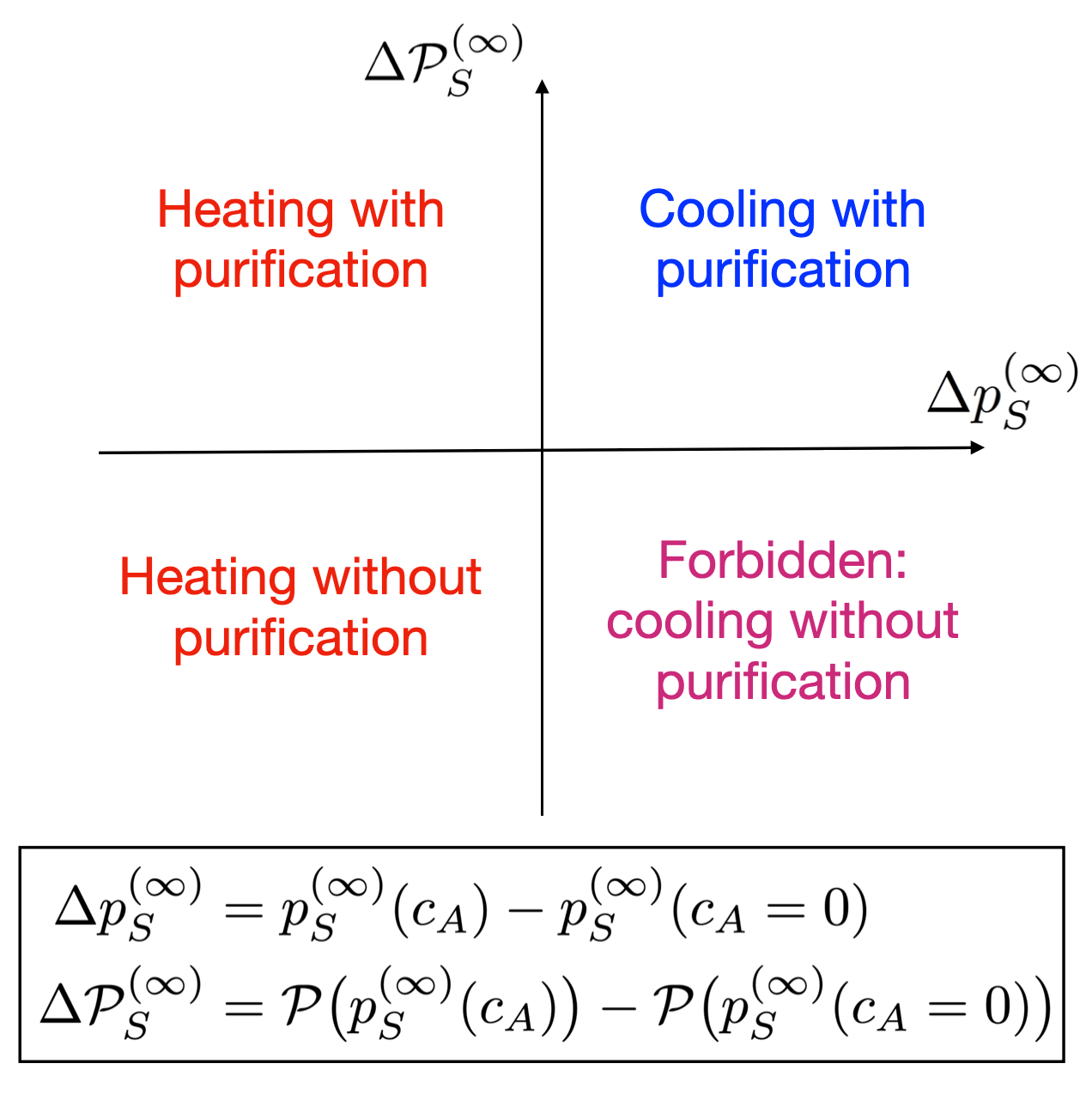}
    \caption{Summary of the different operating regimes of the coherent environment in relation to cooling and purification.}
    \label{fig:CoolP}
\end{figure}


Importantly, the flip-flop interaction is energy conserving, so that no external work is performed during the collisions. Cooling below the bath temperature must therefore originate from a nonthermal resource supplied by the ancillas. In the present model, this resource is quantum coherence.  
This aspect will be described in Sec. \ref{sec:thermodynamics}.

In general, Fig. \ref{fig:CoherentAncillaImpactOnSteadyState} reveals that the impact of environmental coherence is not universal, but is controlled by a competition between the qubit frequency $\omega$ and the interaction strength $J$, providing a route to harness environmental coherence for the preparation and control of the quantum state.
Figure \ref{fig:CoolP} summarizes the different operating regimes of the coherent environment in our model in relation to cooling and purification.

In Appendix \ref{app:number_of_iterations_to_converge}, we analytically derive an upper bound on the number of RI steps required to prepare the system from a given initial state to a desired target state using the Markovian RI protocol. In particular, we examine whether environmental coherence accelerates or slows down the relaxation dynamics. We observe that augmenting the ancilla with coherence typically increases the number of iterations required for the system to reach its steady state.

In Appendix~\ref{sec:QC}, we further investigate the extent to which coherence is retained in the system in steady state and can thus serve as a quantum resource. We characterize this steady-state coherence as a function of the coherence initially present in the environment, using two standard coherence measures from the resource theory of quantum coherence: the \textit{relative entropy of coherence} and the \textit{$l_1$-norm of coherence} \cite{Plenio2014}. 

\subsection{Preparing thermal states by coherent ancillas}

By tuning the ancillary state and the parameters $J$ and $\omega$, the system can be driven toward a desired steady-state target that is nondiagonal in the energy eigenbasis. For applications requiring states that are diagonal in the energy eigenbasis, such as thermal-state preparation (see, e.g., Ref.~\citenum{Chen2023}), the residual steady-state coherence must subsequently be removed without altering the populations. This can be achieved by coupling the system to a purely dephasing environment.
We consider a purely dephasing CPTP map defined as
\begin{equation}
    \Phi_{Z}(\bullet) = \text{Tr}_A(\hat U_Z\bullet\otimes\rho_A\hat U_Z^\dagger),
    \label{eq:dephasingCPTP}
\end{equation}
with
\begin{equation}
    \hat U=e^{-i\hat H_Z\tau}.
\end{equation}
Here, $\hat H_Z$ represents the Hamiltonian that generates the purely dephasing dynamics. In particular, $\hat H_Z  =\hat{H}_0+\hat H_I$, with $\hat H_0$ given by Eq. (\ref{eq:totalfreeHamiltonian}) and $\hat H_I=J_{zz}\sigma_z^{S}\otimes\sigma_z^{A}$.

The protocol proceeds as follows. Starting from an initial state $\rho_S^{(0)}$, we evolve the system with the RI protocol as described in Sec. \ref{sec:ModelB} using coherent ancillas until the system reaches a coherent steady state with the desired populations, corresponding to the target temperature. This state is denoted by $\rho_S^{(\infty)}$. 
The advantage of using coherent ancillas, rather than thermal ones, is in the ability to cool the system below the bath's temperature.
We then evolve the system from $\rho_S^{(\infty)}$ under the purely dephasing dynamics described by the map in Eq.~(\ref{eq:dephasingCPTP}) using the ancillary state $\rho_A$, 
eventually reaching a diagonal thermal state for the system, with the desired temperature. Note that at this stage, it does not matter whether the ancilla is coherent or thermal.
Concretely, the CPTP map in Eq.~(\ref{eq:dephasingCPTP}) evolves the system according to 
\begin{align}
    &\tilde p_{S}^{(n+1)}\equiv p_{S}^{(\infty)},\\
    &\tilde c_{S}^{(n+1)} = \tilde c_S^{(n)}\Big[p_A e^{-i(2J_{zz}-\omega)\tau}+(1-p_A)e^{i(2J_{zz}+\omega)\tau)}\Big]\nonumber\\
    &\:\:\:\:\:\:\:\:\:\:\:\:=(c_S^{(\infty)})\Big[p_A e^{-i(2J_{zz}-\omega)\tau}+(1-p_A)e^{i(2J_{zz}+\omega)\tau}\Big]^{n+1}.
    \label{eq:Jzzmap}
\end{align}
Here, we denote by $\tilde p_S^{(n)}$ and $\tilde c_S^{(n)}$ the population of the ground-state and the coherence of the system during purely dephasing dynamics, respectively. Since we take  $\rho_S^{(\infty)}$ as the initial state of the dephasing dynamics, we identify
$\tilde p_S^{(0)} \equiv p_S^{(\infty)}$ and 
$\tilde c_S^{(0)} \equiv c_S^{(\infty)}$. 

Focusing on off-diagonal elements, we define $\lambda=
p_A e^{-i2J_{zz}\tau}+(1-p_A)e^{i2J_{zz}\tau}$, where
after $n$ collisions,
$|\tilde c_{S}^{(n)}|^2=|\lambda|^{2n}|\tilde c_{S}^{(0)}|^2$.
We note  that
$|\lambda|^2=1-4p_A(1-p_A)\sin^2 (2J_{zz}\tau)$, which is smaller than 1 since $0<p_A<1$, and
as long as $\sin (2J_{zz}\tau) \neq 0$, proving that coherence decays to zero during this stage.
Equation (\ref{eq:Jzzmap}) therefore describes pure dephasing: the coherence progressively decays with each collision, while the populations, and hence the population-defined effective temperature, remain unchanged.

This two-stage protocol provides a simple route to thermal-state preparation on digital quantum hardware, extending previous work, Ref.~\citenum{Prositto2025Dynamics}, to a broader range of target temperatures, including temperatures {\it below} that of the bath, as well as to states of higher purity. The approach separates population preparation from coherence removal. In the first stage, coherent ancillas engineer the desired populations, thereby setting the effective
temperature. In the second stage, purely dephasing dynamics removes the residual coherence without altering the populations, yielding a diagonal thermal state at the same effective temperature.

\begin{figure*}[ht!]
    \centering
    \includegraphics[width=1\linewidth]{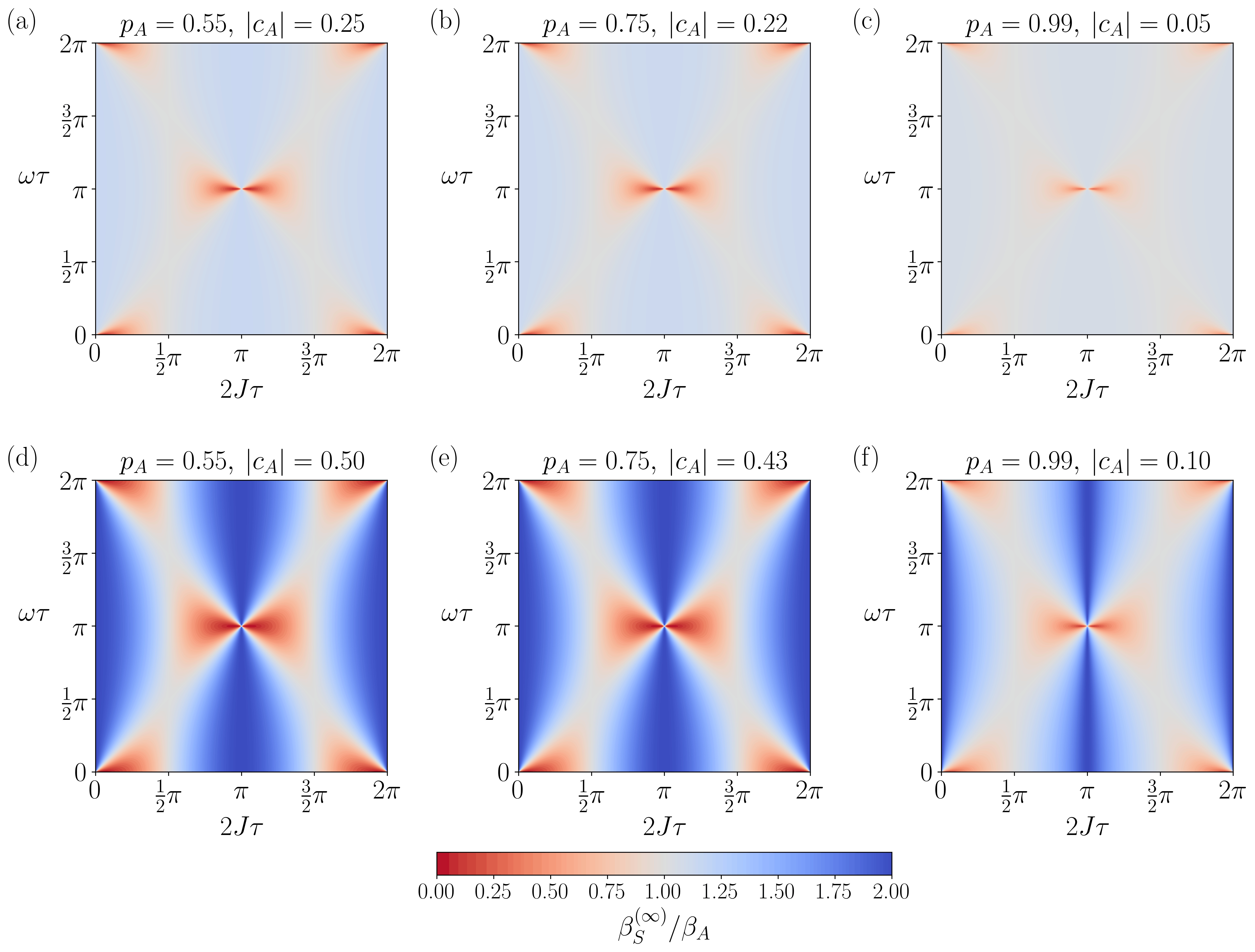}
    \caption{Ratio $\beta_S^{(\infty)}/\beta_A$ as a function of $2J\tau$ and $\omega\tau$, reported for different values of $p_A$ and $|c_A|$. We use Eq. (\ref{eq:coherentancilla_exact_pop_ss}), and apply the definition in Eq. (\ref{eq:beta_definition}). 
    Blue regions correspond to cooling below the bath temperature.
    }
    \label{fig:Temp_map}
\end{figure*}

A natural question that arises is: what is the lowest temperature that can be reached for the system using the protocol outlined here? 
We define the inverse temperature of the steady state system as
\begin{align}
    \beta_S^{(\infty)} = \frac{1}{\omega}\ln\Bigg[\frac{p_S^{(\infty)}}{1-p_S^{(\infty)}}\Bigg]
 =
    \frac{1}{\omega}\ln\Bigg[\frac{2p_A-\Lambda }{2(1-p_A)-\Lambda}\Bigg]
    \label{eq:beta_definition}
\end{align}
which should be compared to the ancilla's inverse temperature,
\begin{align}
    \beta_A = \frac{1}{\omega}\ln\Bigg[\frac{p_A}{1-p_A}\Bigg].
\end{align}
It can be shown that
\begin{equation}
\frac{\beta_S^{(\infty)}}{\beta_A}
= \frac{
\operatorname{arctanh}\left(\dfrac{2p_A-1}{1-\Lambda}\right)
}{ \operatorname{arctanh}(2p_A-1)
}.
\label{eq:difftemp}
\end{equation}
Although alternative definitions of effective temperature have been proposed, e.g., based on exchanged heat \cite{Man23} or system correlations \cite{Lu2026}, here we adopt the standard definition based on the relative  populations of the ground and excited states. This choice is natural for a two-level system, since our protocol transforms the coherent steady state into a thermal state without modifying its populations. Moreover, our system has no internal structure that can support correlations.

In Figure \ref{fig:Temp_map}, we report the relative temperature of the system with respect to the ancilla, presented as a function of $2 J \tau$ and $\omega\tau$ for different choices of ancillas population and coherence, $p_A$ and $|c_A|$. 
The regions with $\beta_S^{(\infty)}/\beta_A>1$ correspond to cooling below bath temperature, while $\beta_S^{(\infty)}/\beta_A<1$ corresponds to heating above bath temperature. Consistently with the cooling condition in Eq.~(\ref{eq:coolingcond}), the effect of coherence strongly depends on the interplay between the interaction scale $J$ and the qubit frequency $\omega$, producing distinct regions of cooling and heating. Increasing $|c_A|$ enhances the deviation of the system temperature from
the bath temperature, demonstrating that environmental coherence provides a flexible resource to control the target temperature. The effect becomes less pronounced as $p_A$ approaches unity, since positivity restricts the maximum allowed coherence to $|c_A|\leq\sqrt{p_A(1-p_A)}$, which vanishes as $p_A\rightarrow1$.


\section{Thermodynamics: coherence as a resource for anomalous heat flow}
\label{sec:thermodynamics}

In this section, we discuss the thermodynamics of our model and, in particular, how environmental coherence modifies the energy balance and hence heat transfer between the system and ancilla. 

Since the flip-flop interaction is energy conserving, no work is performed during the dynamics. For completeness, we recall that, in RI models, the work associated with a single interaction step is defined as \cite{Barra2015,Prositto2025Dynamics}
\begin{equation}
    W^{(n+1)} = \text{Tr}\left[\left(\hat U(\tau)^\dagger\hat H_I\hat U(\tau)-\hat H_I\right)\rho_S^{(n)}\otimes\rho_A\right].
\end{equation}
It can be interpreted as the energy required to compensate for the change in the interaction energy after coupling and subsequent decoupling of the system and the ancilla. In the present model $[\hat H_I,\hat U(\tau)]=0$, thus $W^{(n)}=0$. Since energy exchange is provided in the form of heat, we turn our focus to the magnitude and direction of the heat transfer.

The heat exchanged between the system and the ancilla during a single interaction step is defined as \cite{Barra2015,Prositto2025Dynamics}
\begin{align}
    Q&^{(n+1)\nonumber}\\ &= \text{Tr}\left[\left(\hat U(\tau)^\dagger\unit_S\otimes \hat H_A\hat U(\tau)-\unit_S\otimes\hat H_A\right)\rho_S^{(n)}\otimes\rho_A\right].
    \label{eq:heat_def}
\end{align}
A positive value corresponds to heat flow towards the ancilla. Evaluating this expression at steady state using Eq. (\ref{eq:coherentancilla_exact_pop_ss}) and Eq. (\ref{eq:coherentancilla_exact_coh_ss}), it can be shown that $Q^{(\infty)}=0$. Thus, in steady state, there are no heat or work currents.

At intermediate steps, similarly to Ref. \citenum{DeChiara2022}, we decompose the initial state of each ancilla into the sum of two contributions, one diagonal and one purely off-diagonal,  $\rho_A=\rho_A^{\text{(diag)}}+\rho_A^{\text{(off-diag)}}$ with
\begin{align}
    \rho_A^{\text{(diag)}} = \begin{pmatrix}
        p_A&0\\0&1-p_A
    \end{pmatrix}\:\:\text{and}\:\:\rho_A^{\text{(off-diag)}} = \begin{pmatrix}
        0&c_A\\c_A^*&0
    \end{pmatrix}.
\end{align}
This decomposition allows us to identify the contributions arising from the ancilla populations and from its coherence,
\begin{align}
    Q^{(n+1)} = Q^{(n+1)}_{inc} + Q^{(n+1)}_{coh},
\end{align}
with
\begin{align}
    &Q_{inc}^{(n+1)} = \omega(p_A-p_S^{(n)})\sin^2\left(2J\tau\right)\label{eq:Qinc},\\[0.3 em]
    &Q_{coh}^{(n+1)} = -\omega\sin(4J\tau)Im\left(c_S^{(n)}c_A^*\right).
\label{eq:Qcoh}
\end{align}
Here, the term $Q^{(n+1)}_{inc}$ denotes the contribution arising from the diagonal (incoherent) part of the ancilla state, while $Q^{(n+1)}_{coh}$ originates from its off-diagonal (coherent) part, and thus captures the contribution of environmental coherence. 

The contribution $Q^{(n+1)}_{inc}$ corresponds to the heat transfer that arises from the imbalance of the population, i.e., the difference in the effective temperatures between the system and the ancilla. It represents the conventional, coherence-independent contribution to heat transfer, which tends to reduce the population imbalance, driving the system toward equilibrium with the environment. Consequently, its sign, and therefore the direction of the associated heat flow, is determined solely by the population imbalance: When $p_A-p_S^{(n)}>0$, the ancilla is colder than the system, and heat flows from the system toward the ancilla. 

In contrast, $Q^{(n+1)}_{coh}$, the contribution due to environmental coherence, has a more complex behavior: its sign is determined not only by the coherence of the system and ancilla, but also by the interaction strength $J$ and the interaction time $\tau$. Consequently, environmental coherence can either enhance or suppress energy transfer associated with the population imbalance and, under suitable conditions, even reverse its direction. The coherent contribution can allow heat to flow against the direction dictated by the population imbalance, thereby allowing the system to be cooled below the bath temperature. This reversal in the heat flow direction is termed an \textit{anomalous heat flow}. 

Anomalous heat flow has been extensively investigated, both theoretically \cite{Bera2017,Rudolph2012,Partovi2008,Rudolph2010,Amaral2025} and 
experimentally \cite{Micadei2019,Lu2026}. Recently,
Ref.~\citenum{Lu2026} experimentally demonstrated anomalous heat flow in a repeated-interaction setup implemented using nuclear magnetic resonance (NMR), with a molecular system providing controllable quantum coherence. In that experiment, anomalous heat flow arose from quantum correlations between different constituents of the molecular system, whereas the individual ancillas were prepared in thermal, uncorrelated states. Thus, the internal structure and correlations of the system played a central role in the mechanism underlying the observed anomalous heat flow.

Our model reveals a different mechanism. We show that anomalous heat flow can arise even for a single-qubit system, without internal structure or correlations within the system. Instead, the effect originates from quantum coherence in the ancillas. In particular, the coherent contribution to the 
heat transfer can be nonzero even when the system and ancilla have identical populations in the energy basis, for which the conventional population-driven contribution vanishes: When $p_S^{(n)}=p_A$, $Q_{\mathrm{inc}}^{(n+1)}=0$, while $Q_{\mathrm{coh}}^{(n+1)}$ can be nonzero in the presence of system and ancilla coherences. 

\begin{figure} 
    \centering
    \includegraphics[width=1\linewidth]{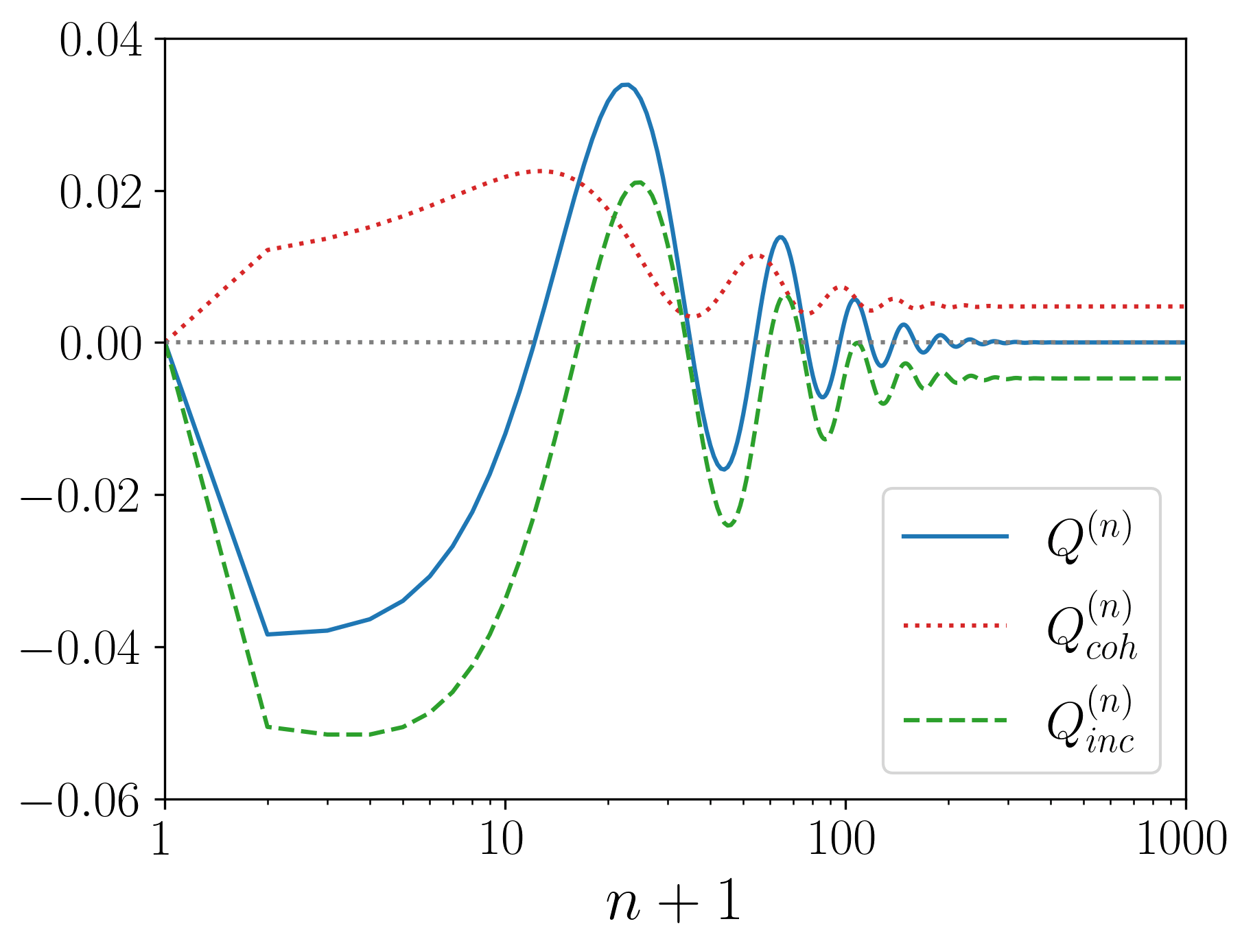}
    \caption{Instance of heat dynamics. We compute the total heat, Eq.~(\ref{eq:heat_def}), and its components, Eq.~(\ref{eq:Qinc}) and Eq.~(\ref{eq:Qcoh}), along a trajectory. The initial state of the system and the ancilla are chosen at random, $p_S^{(0)}=0.58$, $c_S^{(0)} =-0.12-0.46i$ and $p_A=0.89$, $c_A=-0.28-0.00032i$, respectively. We set  $J=1$, $\omega=1$, and $\tau=0.1$.}
    \label{fig:CoherentAncilla_InstanceHeatDynamics}
\end{figure}

\begin{figure*}[ht!]
    \centering
    \includegraphics[width=1\linewidth]{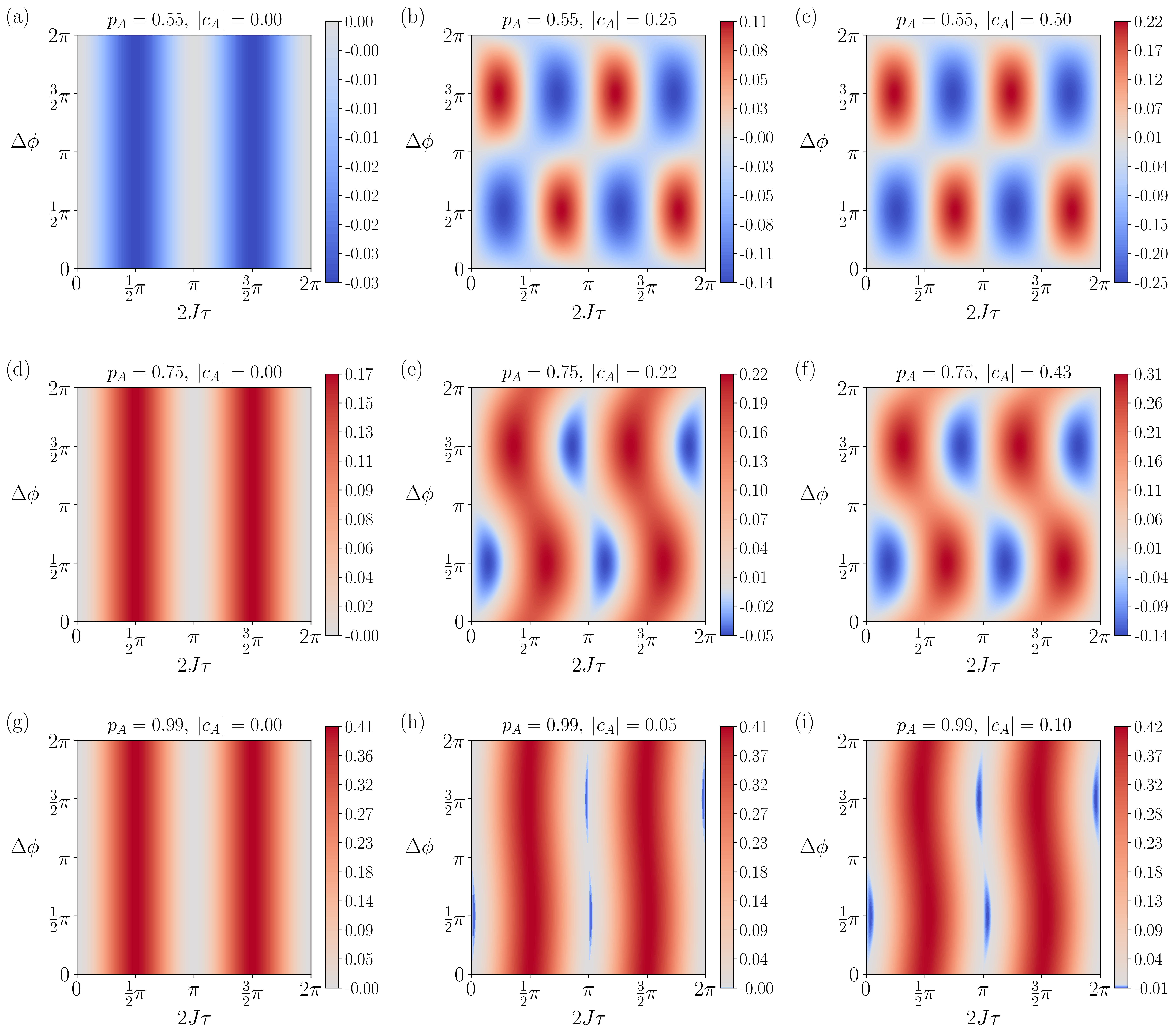}
    \caption{Heat exchanged in a single RI step between the system and the ancilla as a function of $2J\tau$ and $\Delta\phi$, i.e. the relative phase between the coherence of the system and of the ancilla. We calculated the heat as the sum of Eq.~(\ref{eq:Qinc}) and Eq.~(\ref{eq:Qcoh_phase}). We chose the initial state of the system at random,  $p_S^{(0)}=0.58$, $c_S^{(0)} =-0.12-0.46i$, and set $\omega=1$. We computed the heat one step after the initial state, indicating with a positive sign (red) the heat deposited on the ancilla.}
    \label{fig:TotalHeatMap}
\end{figure*}


Expressing the system and ancilla coherences in a polar form, we  write Eq. (\ref{eq:Qcoh}) as
\begin{equation}
    Q_{coh}^{(n+1)} =-\omega|c_S^{(n)}||c_A|\sin(4J\tau) \sin(\Delta\phi^{(n)}),
    \label{eq:Qcoh_phase}
\end{equation}
where $\Delta\phi^{(n)}=\phi_S^{(n)}-\phi_A$ with $c_S^{(n)}=|c_S^{(n)}|e^{i\phi^{(n)}_S}$ and $c_A=|c_A|e^{i\phi_A}$. This shows that, in determining the direction of heat flow, the relative phase between the system and ancilla coherences plays a key role, while their magnitudes only determine the magnitude of the heat transferred. 

In Fig.~\ref{fig:CoherentAncilla_InstanceHeatDynamics}, we illustrate this behavior along a representative trajectory, showing total heat transfer at each step, Eq.~(\ref{eq:heat_def}), together with its coherent and incoherent contributions, Eqs.~(\ref{eq:Qcoh}) and (\ref{eq:Qinc}), respectively.

Figure \ref{fig:TotalHeatMap} shows the total heat exchanged during a representative RI step between the system and the ancilla, as defined in Eq.~(\ref{eq:heat_def}), as a function of the dimensionless coupling $2J\tau$ and the phase difference $\Delta\phi$ between the coherence of the system and the ancilla. We omit the step index for simplicity. Positive (negative) values indicate heat flow from the system to the ancilla (from the ancilla to the system).

For a fixed initial state of the system, we monitor the heat exchanged in the subsequent RI step. In the absence of environmental coherence, $c_A=0$, the coherent contribution in Eq.~(\ref{eq:Qcoh}) vanishes, leaving only the incoherent contribution arising from the population, or temperature, imbalance, Eq.~(\ref{eq:Qinc}). Accordingly, heat flows from hot to cold, as expected from the second law of thermodynamics, see Fig. \ref{fig:TotalHeatMap} (a), (d), (g).

When coherence is present in the environment, anomalous heat flow can emerge, with heat flowing from cold to hot, seemingly in violation of the second law. However, this apparent violation is enabled by the consumption of environmental coherence as a thermodynamic resource, see Fig. \ref{fig:TotalHeatMap} (b), (c), (e), (f), (h), (i). The phase difference $\Delta\phi$ determines the direction of the heat flow, while the magnitude of the heat is controlled by the magnitude of coherence in the environment. 

As expected, the larger the population imbalance between the system and the ancilla, the smaller the region of parameter space supporting anomalous heat flow. This is because the incoherent contribution to heat, Eq.~(\ref{eq:Qinc}), which drives heat transfer toward thermal equilibration, becomes increasingly dominant. This suggests that for practical applications, also the fine-tuning of the ancilla temperature (i.e. populations), in addition to its coherence, can be important for achieving a certain task.

The identification of ancillary coherence as a thermodynamic resource can also be used to study aspects of nonequilibrium heat transfer between two baths, mediated by the system. This would generalize \cite{Gabrielle2026}, as well as studies exploring the nonequilibrium spin-boson model; for a partial list of references, see \cite{Segal05,nazim14,Nick21,Gelb26}. When the baths have different temperatures and nonzero coherences, the direction of the heat current, its magnitude, and the thermal rectification effect could be all controlled by baths' coherence.

\section{Conclusions}
\label{sec:conclusions}

We investigated the relaxation dynamics and steady-state properties of a qubit interacting sequentially with coherent ancillas within a Markovian repeated-interaction framework. For a minimal model with an energy-conserving flip-flop interaction, we derived an exact analytical solution without assuming weak environmental coherence or weak system-ancilla coupling. 

We summarize the main impact of our work:

1. Dissipative state preparation: We show that preparing ancillas with coherence enlarges the accessible steady-state manifold for the system, yet without changing the basic collision architecture.
Fundamentally, coherent ancillas break the symmetry responsible for population–coherence decoupling. Thus, bath coherence allows access to states unavailable to an incoherent reservoir.

2. Cooling beyond conventional thermalization: 
We prove that the system can reach ground-state population {\it higher} than that of each incoming ancilla, corresponding to a cooling process of the system (see e.g., Fig. \ref{fig:CoherentAncillaImpactOnSteadyState}(g)).
We point out that cooling implies enhanced steady-state purity relative to the thermal-ancilla protocol, but purification need not imply cooling (Fig. \ref{fig:CoherentAncillaImpactOnSteadyState}(d)-(f)), blue line). 
These results demonstrate that environmental coherence provides a resource for controlling both the effective temperature and the purity of the system.

3. Coherence-assisted control of heat flow:
We further analyze the thermodynamics underlying these effects. Since the flip-flop interaction is energy conserving, the dynamics involves no work, allowing the role of environmental coherence in energy exchange to be isolated. We decompose the heat exchanged in each collision into population-driven and coherence-driven contributions and show that environmental coherence can modify, and even reverse, the direction of heat flow. Remarkably, heat transfer can occur even when the system and ancilla have identical populations, and hence in the absence of the conventional population imbalance driving heat exchange. For fixed system-ancilla coupling parameters, the direction of the coherence-induced heat flow is controlled by the relative phase between the system and ancilla coherences, providing a minimal mechanism for anomalous heat flow that does not require a multipartite system or internal correlations.

These results could have an impact on the field of autonomous quantum error correction and stabilization \cite{ChenW}: population-coherence coupling may be used to stabilize logical states and engineer a nontrivial steady state that is not achievable with thermal ancillas.
Concretely, our results suggest a route toward dissipative state preparation on digital quantum hardware using coherent environmental resources. By tuning the ancillary state and the Hamiltonian parameters, the RI dynamics can first prepare a coherent steady state with the desired populations and effective temperature; a subsequent purely-dephasing stage can then remove the residual coherence without changing these populations, yielding a diagonal thermal state. This two-stage protocol extends the range of thermal states accessible through RI dynamics, including states colder than the environmental ancillas. 
 
While the present analysis focuses on a single qubit, the underlying mechanism is not restricted to two-level systems. Extending coherence-assisted repeated-interaction protocols to multilevel and many-body systems could provide a route to dissipative state preparation, enabling the engineering of nonthermal population distributions, quantum coherences and correlations, and potentially entangled steady states.

\begin{acknowledgments}
The authors acknowledge the fruitful discussion with Barbara Amaral.
A.P. was supported by the Department of Physics at University of Toronto and the research project: ``Quantum Software Consortium: Exploring Distributed Quantum Solutions for Canada" (QSC). QSC is financed under the National Sciences and Engineering Research Council of Canada (NSERC) Alliance Consortia Quantum Grants \#ALLRP587590-23. D.S. acknowledges the NSERC Discovery Grant.
\end{acknowledgments}

\bibliography{references}

\appendix
\begin{widetext}

\section{Population-coherence decoupling in repeated-interaction dynamics}
\label{app:population_coherence}

In this Appendix, we establish a general criterion for the decoupling of population and coherence dynamics in an RI model. Importantly, population–coherence decoupling is a property of the reduced quantum channel acting on the system and is ultimately rooted in the symmetries of the underlying microscopic model. This provides a direct connection between the microscopic structure of the RI model and the resulting population and coherence dynamics.

Consider the system dynamics generated in a single interaction by the quantum channel 
\begin{equation}
\Phi(\rho_S) = \text{Tr}_A\left[\hat U(\rho_S\otimes\rho_A)\hat U^\dagger    \right],
    \label{eq:app_channel}
\end{equation}
where $\hat U=e^{-i\hat H\tau}$ is the time evolution operator and $\rho_A$ is the state of the incoming ancilla.

Let $\{|n\rangle_S\}$ denote the reference basis in which the system populations and coherence are defined. In the present work, this corresponds to the energy eigenbasis of the system Hamiltonian $\hat H_S$, but it generally can be the eigenbasis of a generic operator $\hat Z_S$ with support on the Hilbert space of the system.  
We introduce the (system) dephasing map in this basis,
\begin{equation}
    \mathcal{D}_S(\hat X_S) = \sum_{n} |n\rangle_S\langle n| \hat X_S   |n\rangle_S\langle n|,
    \label{eq:app_dephasing}
\end{equation}
where $\hat X_S$ represents a generic operator acting in the Hilbert space of the system.

Similarly, we can define $\mathcal{D}_A$, the dephasing map for the ancilla, with respective populations and coherence defined in the basis $\{|n\rangle_A\}$, eigenbasis of a certain operator $\hat Z_A$ acting on the Hilbert space of the ancilla.

The operator space of the system can then be decomposed into a diagonal
sector and an off-diagonal sector,
\begin{equation}
    \hat X_S=\hat X_{S,\rm diag}+\hat X_{S,\rm coh},
\end{equation}
with
\begin{equation}
  \hat   X_{S,\rm diag}=\mathcal D_S(\hat X_S),
    \qquad
\hat     X_{S,\rm coh}=(\mathcal I-\mathcal D_S)(\hat X_S),
\end{equation}
where $\mathcal I$ is the (system) identity map. 

Population and coherence dynamics are decoupled when the channel preserves
these two sectors separately: diagonal operators are mapped onto diagonal operators, while off-diagonal operators are mapped onto off-diagonal operators. This condition is equivalent to
\bea
\mathcal D_S[\Phi(\hat X_S)] =  \Phi [\mathcal D_S(\hat X_S)].
\label{eq:app_dephasing_covariance}
\eea
Suppose first that Eq.~(\ref{eq:app_dephasing_covariance}) holds. For a diagonal operator $\hat X_{S,\rm diag}$,
$\mathcal D_S(\hat X_{S,\rm diag})=\hat X_{S,\rm diag}$, and therefore
\begin{equation}
    \mathcal D_S[\Phi(\hat X_{S,\rm diag})] =
    \Phi[\mathcal D_S(\hat X_{S,\rm diag})] =
    \Phi(\hat X_{S,\rm diag}),
\end{equation}
showing that $\Phi(\hat X_{S,\rm diag})$ is diagonal. Similarly, for an
off-diagonal operator $\hat X_{S,\rm coh}$,
$\mathcal D_S(\hat X_{S,\rm coh})=0$, such that
\begin{equation}
    \mathcal D_S[\Phi(\hat X_{S,\rm coh})] =
    \Phi[\mathcal D_S(\hat X_{S,\rm coh})]=0.
\end{equation}
Thus, $\Phi(\hat X_{S,\rm coh})$ does not contain a diagonal component. Conversely, if the diagonal and off-diagonal sectors are separately invariant under
$\Phi$, then for an arbitrary operator
$\hat X_S=\hat X_{S,\rm diag}+\hat X_{S,\rm coh}$,
\bea
    \mathcal D_S[\Phi(\hat X_S)]
    &=& \mathcal D_S[\Phi(\hat X_{S,\rm diag})] +
    \mathcal D_S[\Phi(\hat X_{S,\rm coh})]
    \nonumber\\
    &=&    \Phi(\hat X_{S,\rm diag}) =
    \Phi[\mathcal D_S(\hat X_S)],
\eea
which proves Eq.~(\ref{eq:app_dephasing_covariance}). See also Refs.~\citenum{Spekkens2016,Gu2020,Rudolph2015}.

Therefore,  Eq.~(\ref{eq:app_dephasing_covariance})  provides a {\it necessary and sufficient } criterion for population--coherence decoupling. Notice that
this criterion is formulated directly in terms of the reduced channel
$\Phi$. However, Eq.~(\ref{eq:app_dephasing_covariance}) cannot, in general, be replaced by a necessary and sufficient condition on the total Hamiltonian alone, since the partial trace over the ancilla and the ancilla state $\rho_A$  can generate symmetries of the reduced dynamics that are not symmetries of the full system-ancilla unitary evolution.

We next provide a  microscopic {\it sufficient} condition for
Eq.~(\ref{eq:app_dephasing_covariance}). There are two possible strategies.

\subsection{Strategy 1: Representing the dephasing map as a time average}
It is possible to demonstrate that when the dynamics is unitary, the dephasing map $\mathcal{D}_S$ admits the following integral representation,
\begin{align}
        \mathcal{D}_S(\rho_S) = \lim_{T\rightarrow +\infty} \frac{1}{2T}\int_{-T}^{T}e^{-i\hat Z_St}\rho_S e^{i\hat Z_St} dt,
        \label{eq:D_int_rep}
\end{align}
for any density matrix $\rho_S$  of the system, see Sec. \ref{sec:deph}. Here, $\hat Z_S$ is a linear self-adjoint operator with support on the Hilbert space of the system that defines the relevant orthonormal basis for this representation. In our model, $\hat Z_S\equiv\hat H_S$. 

Therefore, the population-coherence decoupling condition, Eq.~(\ref{eq:app_dephasing_covariance}),  can be written in the basis defined by the operator $\hat Z_S$ as
\begin{align}
    \mathcal{D}_S\Big(\text{Tr}_A\Big[\hat U\rho_S\otimes\rho_A \hat U^\dagger\Big]\Big) &=\text{Tr}_A\Big[\hat U\mathcal{D}_S(\rho_S)\otimes\rho_A \hat U^\dagger\Big] 
   \\
   \lim_{T\rightarrow+\infty}\frac{1}{2T}\text{Tr}_A\Bigg[\int_{-T}^{T}e^{-i\hat Z_St}\otimes\unit_A \hat U\rho_S\otimes\rho_A \hat U^\dagger e^{i\hat Z_St}\otimes\unit_A\Bigg]dt&=\lim_{T\rightarrow+\infty}\frac{1}{2T}\text{Tr}_A\Bigg[\int_{-T}^{T}\hat Ue^{-i\hat Z_S t}\otimes\unit_A\rho_{S}\otimes\rho_A e^{i\hat Z_S t}\otimes\unit_A\hat U^\dagger dt\Bigg],
   \label{eq:dec_int_cond}
\end{align}
for any density matrix $\rho_S$ ($\rho_A$) of the system (ancilla).

A {\it sufficient but not necessary} condition to decoupling satisfying Eq.~(\ref{eq:dec_int_cond}) is that 
\begin{equation}
[\hat H, \hat Z_S \otimes \unit_A] = 0.
\label{eq:HZS}
\end{equation}
We emphasize that if this Equation does not hold, coupling of population to coherence may or may not occur - we cannot conclude anything
since Eq. (\ref{eq:HZS}) is a sufficient but not necessary condition for decoupling.

Indeed, although the flip-flop model does not satisfy Eq.~(\ref{eq:HZS}), population and coherence dynamics remain decoupled for diagonal ancillas as a consequence of the joint symmetry below, in Eq.~(\ref{eq:joint}).

Similarly, in case of diagonal ancillas, $\mathcal D_A (\rho_A)=\rho_A$, we get as a {\it sufficient but not necessary} condition for population-coherence decoupling
\bea
[\hat H, \hat Z_S \otimes \unit_A+\unit_S\otimes\hat Z_A] = 0.
\label{eq:joint}
\eea

In case of our flip-flop model, we have $[\hat H, \hat Z_S \otimes \unit_A+\unit_S\otimes\hat Z_A] = 0$, since the total Hamiltonian is energy conserving, but $[\hat H, \hat Z_S \otimes \unit_A] \neq 0$. This is why when the ancilla is diagonal we observe decoupling, while in our model where the ancilla is coherent, the symmetry leading to population-coherence decoupling breaks.

\subsection{Strategy 2: Phase rotation covariance}

Let the populations and coherence of the system be defined in the eigenbasis of a Hermitian operator $\hat Z_S$, and introduce the phase rotation operator
\begin{equation}
   \hat  V_S(\phi)=e^{-i\phi \hat Z_S}.
\end{equation}
Similarly, let $\hat Z_A$ define the corresponding reference basis of the
ancilla, with
\begin{equation}
    \hat V_A(\phi)=e^{-i\phi \hat Z_A}.
\end{equation}
Suppose that the joint unitary commutes with the simultaneous phase
rotation,
\begin{equation}
    [\hat U,\hat V_S(\phi)\otimes \hat V_A(\phi)]=0
    \qquad
    \forall\,\phi,
    \label{eq:app_joint_covariance}
\end{equation}
or, equivalently,
\begin{equation}
    [\hat H,\hat Z_S\otimes\mathbbm 1_A+
       \mathbbm 1_S\otimes \hat Z_A]=0.
    \label{eq:app_conservation}
\end{equation}
Furthermore, suppose that the incoming ancilla is invariant under its
corresponding phase rotations,
\begin{equation}
    \hat V_A(\phi)\rho_A \hat V_A^\dagger(\phi)=\rho_A
    \qquad
    \forall\,\phi,
    \label{eq:app_ancilla_invariant}
\end{equation}
equivalent to
\begin{equation}
    [\rho_A,\hat Z_A]=0,
\end{equation}
as long as $\hat Z_A$ is  nondegenerate.
Under these conditions, the reduced system channel is phase covariant:
\begin{equation}
    \Phi\left[
    \hat V_S(\phi)\rho_S \hat V_S^\dagger(\phi)
    \right]  =\hat V_S(\phi)\Phi(\rho_S)\hat V_S^\dagger(\phi).
    \label{eq:app_phase_covariance}
\end{equation}
To see this, we write
\bea
\Phi\left[
\hat V_S(\phi)\rho_S \hat V_S^\dagger(\phi)
\right]
={\rm Tr}_A\left\{\hat  U \left[ \hat V_S(\phi)\rho_S \hat V_S^\dagger(\phi) \otimes\rho_A \right] \hat U^\dagger
\right\}.
\eea
Using Eq. (\ref{eq:app_ancilla_invariant}), this becomes
\begin{align}
&{\rm Tr}_A\left\{
\hat U
\left[\hat V_S(\phi)\otimes \hat V_A(\phi)\right]
(\rho_S\otimes\rho_A)
\left[\hat V_S^\dagger(\phi)\otimes \hat V_A^\dagger(\phi)\right]
\hat U^\dagger
\right\}.
\end{align}
Applying Eq.~(\ref{eq:app_joint_covariance}) and using the invariance of
the partial trace under unitary transformations acting only on the
ancilla gives Eq.~(\ref{eq:app_phase_covariance}).
The dephasing operation can be obtained by averaging these phase rotations,
using the time-average representation introduced in Eq.~(\ref{eq:D_int_rep}),
\begin{equation}
    \mathcal D_S(\hat X_S)
    =
    \lim_{T\rightarrow\infty}\frac{1}{2T}    \int_{-T}^{T} \hat V_S(\phi)\hat X_S \hat V_S^\dagger(\phi)\,d\phi,
    \label{eq:app_dephasing_average}
\end{equation}
provided that the spectrum of $\hat Z_S$ is chosen such that the averaging
removes all off-diagonal matrix elements in the reference basis.
Applying Eq.~(\ref{eq:app_phase_covariance}) inside this integral yields
Eq. (\ref{eq:app_dephasing_covariance}).

For the flip-flop interaction considered in this work, Eq.~(\ref{eq:app_conservation}) is satisfied as a consequence of energy conservation. For diagonal ancillas, Eq.~(\ref{eq:app_ancilla_invariant}) is also satisfied, and the reduced dynamics therefore obeys the decoupling condition (\ref{eq:app_dephasing_covariance}). For coherent ancillas, Eq.~(\ref{eq:app_ancilla_invariant}) is violated, so this symmetry argument no longer guarantees decoupling. Importantly, violation of Eq.~(\ref{eq:app_ancilla_invariant}) alone does not imply population-coherence coupling.
\subsection{Time average representation of the dephasing map}
\label{sec:deph}

We show here how the integral representation of the dephasing map, Eq.~(\ref{eq:D_int_rep}) is obtained for a finite-dimensional system.
By the spectral theorem, in the eigenbasis $\{\ket{n}_S\}$ of $\hat Z_S$, we have
\begin{align}
  \hat Z_S &= \sum_{n} \omega_n \ket{n}\bra{n},\\[0.3em]
 \rho_S &= \sum_{m,n}\rho_{S,mn}\ket{m}\bra{n}.   
\end{align}
Therefore,
\begin{align}
    e^{-i\hat Z_St}\rho_S e^{i\hat Z_S t}  = \sum_{m,n=1}^{d}\rho_{S,mn}e^{-i\hat Z_S t}\ket{m}\bra{n}e^{i\hat Z_S t}
    =\sum_{m,n=1}^{d}\rho_{S,mn}e^{-i(\omega_m-\omega_n)}\ket{m}\bra{n}.
    \label{eq:D_int_rep_integrand}
\end{align}
The time integral is
\begin{align}
\lim_{T\rightarrow+\infty}\frac{1}{2T}\int_{-T}^T e^{-i\hat Z_St}\rho_S e^{i\hat Z_S t}dt = \sum_{m,n=1}^{d} \rho_{S,mn}\Bigg(\lim_{T\rightarrow+\infty}\frac{1}{2T}\int_{-T}^T e^{-i(\omega_m-\omega_n)t}dt \Bigg) \ket{m}\bra{n}.  
\label{eq:D_int_rep_integral}
\end{align}
However
, the improper integral enclosed in parentheses in the RHS of Eq.~(\ref{eq:D_int_rep_integral}) is equal to
\begin{equation}
    \lim_{T\rightarrow+\infty}\frac{1}{2T}\int_{-T}^T e^{-i(\omega_m-\omega_n)t}dt =\delta_{mn},
\end{equation}
since
\begin{align}
    1)\text{ if $n=m$:}\nonumber\\
    &\lim_{T\rightarrow+\infty}\frac{1}{2T}\int_{-T}^T e^{-i(\omega_m-\omega_n)t}dt = \lim_{T\rightarrow+\infty}\frac{1}{2T}\int_{-T}^T dt =\lim_{T\rightarrow+\infty}\frac{2T}{2T} = 1\\
    2)\text{ if $n\neq m$:}\nonumber\\
    &\lim_{T\rightarrow+\infty}\frac{1}{2T}\int_{-T}^T e^{-i(\omega_m-\omega_n)t}dt =\lim_{T\rightarrow+\infty}\frac{1}{2T}\Bigg(\frac{e^{i(\omega_n-\omega_m)T}-e^{-i(\omega_n-\omega_m)T}}{i(\omega_n-\omega_m)}\Bigg)\nonumber\\&\:\:\:\:\:\:\:\:\:\:\:\:\:\:\:\:\:\:\:\:\:\:\:\:\:\:\:\:\:\:\:\:\:\:\:\:\:\:\:\:\:\:\:\:\:\:\:\:\:\:\:\:\:\:\:\:\:=\lim_{T\rightarrow+\infty}\frac{1}{2T}\Bigg(\frac{2\sin\left((\omega_n-\omega_m)T\right)}{\omega_n-\omega_m}\Bigg)=0. 
\end{align}
In the last expression, the sine function at the numerator is limited to $[-1,1]$, while the denominator goes to $+\infty$.
We also assumed a nondegenerate spectrum of \(\hat Z_S\).
Therefore,
\begin{equation}
    \lim_{T\rightarrow+\infty}\frac{1}{2T}\int_{-T}^T e^{-i\hat Z_St}\rho_S e^{i\hat Z_St}dt  = \sum_{m,n=1}^{d}\rho_{S,mn}\delta_{mn}\ket{m}\bra{n} = \sum_{m=1}^d\rho_{S,mm}\ket{m}\bra{m},
\end{equation}
which defines a dephasing map.

\subsection{
Population-coherence decoupling with a unique fixed point implies a diagonal steady state, but not vice versa}

Equation~\eqref{eq:app_dephasing_covariance} has an immediate 
implication for the steady state. If the channel $\Phi$ has a 
unique fixed point $\rho_S^{(\infty)}$, population-coherence decoupling 
implies
\begin{align}
\Phi\!\left[\mathcal{D}_S\!\left(\rho_S^{(\infty)}\right)\right]
&=
\mathcal{D}_S\!\left[\Phi\!\left(\rho_S^{(\infty)}\right)\right]
\nonumber\\
&=
\mathcal{D}_S\!\left(\rho_S^{(\infty)}\right).
\end{align}
Hence, $\mathcal{D}_S(\rho_S^{(\infty)})$ is also a fixed point of 
$\Phi$. By uniqueness,
\begin{equation}
\mathcal{D}_S\!\left(\rho_S^{(\infty)}\right)
=
\rho_S^{(\infty)},
\end{equation}
and the steady state is therefore diagonal in the reference basis. 
The converse is not generally true: a channel may possess a diagonal steady state even when populations and coherence are coupled during the transient dynamics.

\section{Differential Equations of Motion}  
\label{app:cohancillas_eom}

Defining $X^{(n)}\equiv\left(p_S^{(n)},c_S^{(n)},\big(c_S^{(n)}\big)^{*}\right)^T$, with the superscript $T$ indicating the operation of matrix transposition, Eq. (\ref{eq:CoherentAncilla_exactsol_pop_nstep}) and Eq. (\ref{eq:CoherentAncilla_exactsol_coh_nstep}) can be written in a vector form
\begin{equation}
    X^{(n+1)} = MX^{(n)} + \Upsilon,
\label{eq:cohancilla_timeevol_vec}
\end{equation}
with
\begin{equation}
    M=
    \begin{bmatrix}
        \cos^2(2J\tau) & \frac{i}{2}\sin(4J\tau) c_A^{*} & -\frac{i}{2}\sin(4J\tau) c_A \\i2c_A\sin(2J\tau) e^{i\omega\tau} &\cos(2J\tau)e^{i\omega\tau} &0\\-i2c_A^{*}\sin(2J\tau)  e^{-i\omega\tau}&0&\cos(2J\tau)e^{-i\omega\tau}
    \end{bmatrix},
    \label{eq:cohancillas_M-matrixevolution}
\end{equation}
and
\begin{equation}
    \Upsilon = 
    \begin{bmatrix}
        p_A\sin^2(2J\tau) \\-ic_A\sin(2J\tau) e^{i\omega\tau}\\i c_A^{*}\sin(2J\tau) e^{-i\omega\tau}
    \end{bmatrix}.
\end{equation}
Equation (\ref{eq:cohancilla_timeevol_vec}) is modified by subtracting $X^{(n)}$ from both sides and Taylor-expanding to second order around $\tau=0$. Dividing by $\tau$, and  subsequently taking the limit $\tau\rightarrow0^+$, we get
\begin{equation}
    \dot{X}(t) = \Lambda X(t) + \bar\Upsilon,
\end{equation}
where
\begin{align}
    \Lambda &\equiv \lim_{\tau\rightarrow0^+} \frac{(M-I)}{\tau} = 
    \begin{bmatrix}
        -2\Gamma &i2J c_A^* &-i2Jc_A\\i4Jc_A&i\omega-\Gamma e^{i\Omega} & 0\\
        -i4c_A^{*}J&0&-i\omega-\Gamma e^{-i\Omega}
    \end{bmatrix},
\end{align}
and
\begin{equation}
    \bar \Upsilon\equiv\lim_{\tau\rightarrow0^+}\frac{\Upsilon}{\tau} = \begin{bmatrix}
        2\Gamma p_A\\-i2c_AJ\\i2c_A^*J
    \end{bmatrix}.
\end{equation}
Here, we set $t = n\tau$ and define $\Gamma \equiv2J^2\tau$ and $\Omega\equiv\omega\tau$, such that they stay constant when $\tau\rightarrow0^+$.
In the case of diagonal ancillas,  $c_A=0$, then
\begin{equation}
   \Lambda=  \begin{bmatrix}
        -2\Gamma&0&0\\0&i\omega-\Gamma e^{i\Omega}&0\\0&0&-i\omega-\Gamma e^{-i\Omega}
    \end{bmatrix},\:\:\:\:\bar \Upsilon = \begin{bmatrix}
        2\Gamma p_A\\0\\0
    \end{bmatrix},
\end{equation}
reducing to the equations of motion derived in Ref.~\citenum{Prositto2025Dynamics}.


\section{Proof that $K\geq0$}
\label{app:K_nonnegative}

We show here that $K\geq0$, see Eq.~(\ref{eq:coherentancilla_ss_denominator}).
Defining $x=\cos(2J\tau)$,  $y=\cos(\omega\tau)$, and $q=4|c_A|^2$,
Eq.~(\ref{eq:coherentancilla_ss_denominator}) is written as
\bea
K = 1+(1-q)x^2-(2-q)xy.
\eea
Positivity of the density matrix implies $|c_A|\leq \frac{1}{2}$, and therefore $0\leq q\leq 1$.
Moreover, since $|y|\leq 1$, $xy\leq |x|$.
Because $2-q\geq 0$, it follows that
\bea
K \geq 1+(1-q)x^2-(2-q)|x|.
\eea
Introducing
$z=|x|$, $0\leq z\leq 1$, we obtain
\bea
K \geq 1+(1-q)z^2-(2-q)z.
\eea
The right-hand side can be factorized as
\bea
1+(1-q)z^2-(2-q)z
=
(1-z)\left[1-(1-q)z\right].
\eea
Since $0\leq z\leq 1$ and $0\leq 1-q\leq 1$, both factors are
nonnegative: $1-z\geq 0$, $1-(1-q)z\geq 0$.
Therefore, $K\geq 0$ for arbitrary values of $J$, $\tau$, and $\omega$, and for every physical ancilla state.

Note that $K=0$ requires $\cos(2J\tau)=\cos(\omega \tau)=\pm1$ with the same sign. At these points, the reduced map is the identity, so the dynamics is frozen and no unique steady state is selected. They are therefore excluded from the relaxing regime considered in the main text.

\section{Proof of $\mathcal{P}(\rho_S^{(\infty)})\leq\mathcal{P}(\rho_A)$ }
\label{app:proof_DeltaPurity}

In this appendix, we prove that within our model the purity of the system in steady state can never exceed the purity of the ancilla, even when environmental coherence is considered, $c_A\neq0$, namely
\bea
\mathcal{P}(\rho_S^{(\infty)})\leq\mathcal{P}(\rho_A).
\eea
We write the purity of the system-qubit at steady state as
\begin{align}
    \mathcal{P}(\rho_S^{(\infty)}) =\frac{1}{2}\left[1+(2p_S^{(\infty)}-1)^2+4|c_S^{(\infty)}|^2\right].
\end{align}
To simplify the notation, we use $x=2p_A-1$, $q=|c_A|^2$, $C=\cos (2 J\tau)$ and $L=\cos (2 J\tau) - \cos (\omega \tau)$. We also reuse previous notation, $D=L^2+\sin^2 (\omega \tau)$, $K=D-4qCL$, see Sec. \ref{sec:ModelB}.
Throughout this proof, we consider $K>0$, for which the steady state is uniquely defined.

Consequently, Eq.~(\ref{eq:coherentancilla_exact_pop_ss}) and Eq.~(\ref{eq:coherentancilla_exact_coh_ss}), which describe the population and coherence of the system in steady state, respectively, take the form
\begin{align}
 2p_S^{(\infty)}-1 &= x + \frac{4qCL}{K}  = x\frac{ D}{K}, \\
| c_S^{(\infty)}|^2 &= \frac{qx^2 D\sin^2 (2J\tau)}{K^2}.
\end{align}
The difference between the steady-state system purity and the ancilla purity is therefore
\bea
\Delta\mathcal{P} \equiv \mathcal{P}(\rho_S^{(\infty)})-
\mathcal{P}(\rho_A)=\frac{x^2}{2K^2}\left[D^2-K^2+4qD\sin^2 (2J\tau) \right] - 2q.
\eea
Algebraic manipulations provide 
\begin{align}
    \Delta\mathcal{P} &= 2q \left(\frac{x^{2}B}{K^2}-1  \right),
    \label{eq:deltapurity}
\end{align}
with $B= D^2-4qC^2L^2$.

It remains to show that \(x^2B\leq K^2\). If \(B\leq0\), this inequality follows immediately since \(K^2>0\). If \(B>0\), positivity of the ancilla density matrix implies $x^2\leq 1-4q$, therefore $x^2B\leq (1-4q)B$.
Therefore, 
\bea
    K^2 - x^2B \geq (D-4q CL)^2 - (1-4q)(D^2-4q  C^2L^2)=4q(D-CL)^2,
\eea
that, being always nonnegative, proves the relation.

In particular, for thermal ancillas, $|c_A|=0$, Eq.~(\ref{eq:deltapurity}) gives $\Delta\mathcal{P}=0$, and therefore $\mathcal{P}(\rho_S^{(\infty)})=\mathcal{P}(\rho_A)$. Thus, the system asymptotically reaches the purity of the ancilla. For coherent ancillas, $|c_A|>0$, this is no longer generally the case: the steady-state purity of the system is strictly smaller than that of the ancilla. Nevertheless, compared with the corresponding thermal ancilla case, environmental coherence can enhance the purity reached by the system in steady state.


\section{Upper bound on the minimum number of collisions required for state preparation}
 \label{app:number_of_iterations_to_converge}

We derive an upper bound on the minimum number of steps required for the system to converge within a prescribed trace-distance threshold $\epsilon$ of the target steady state according to Eq.~(\ref{eq:coherentancilla_exact_pop_ss}) and Eq.~(\ref{eq:coherentancilla_exact_coh_ss}).
In case the target state is a thermal state, this bound will provide an estimation on the minimum number of iterations needed to prepare the desired system populations, i.e., setting the desired temperature; the subsequent application of the pure dephasing map will then completely kill all the coherences leaving a genuine thermal state at the target temperature.

Similarly to Ref.~\citenum{Prositto2025Dynamics}, we choose the trace distance as a metric to determine the distance between the state of the system at a certain step and the target state. 
We use trace distance as a standard operational measure of distinguishability; analogous bounds may be formulated using other distance measures \cite{NielsenChuangBook,WildeBook}. 

The trace distance between two quantum states $\rho$ and $\sigma$ is defined as
\begin{align}
\label{eq:tracedistance}
D&\left(\rho,\sigma\right) = \frac{1}{2} \left| \left| \rho - \sigma \nonumber\right|\right|_{1}=\frac{1}{2}\text{Tr}\Bigg[\sqrt{(\rho-\sigma)^{\dagger}(\rho-\sigma)}\Bigg],
\end{align}
where $||\bullet||_1$ is the \textit{Schatten 1-norm} \cite{NielsenChuangBook,WildeBook}.

Formally, we thus have to find the minimum number of iterations $n^*$ such that
\begin{equation}
    D\left(\rho_S^{(\infty)},\rho_S^{(n^*)}\right)=\frac{1}{2}||\rho_S^{(\infty)}-\rho_S^{(n^*)}||_1<\epsilon
\end{equation}
for a given threshold $\epsilon$.

We start noting that the trace distance (half the sum of the absolute values of the eigenvalues) is
\begin{align}
D\left(\rho_S^{(\infty)},\rho_S^{(k)}\right)&= \sqrt{(p_S^{(k)}-p_S^{(\infty)})^2+|c_S^{(k)}-c_S^{(\infty)}|^2}\nonumber
    \\&= \sqrt{(\delta p_S^{(k)})^2+|\delta c_S^{(k)}|^2}\nonumber
    \\& =\sqrt{(\delta p_S^{(k)})^2+\frac{1}{2}|\delta c_S^{(k)}|^2+\frac{1}{2}|(\delta c_S^{(k)})^*|^2},
\label{eq:tracedistance_explicit}
\end{align}
where we have defined $\delta p_S^{(k)}\equiv p_S^{(k)}-p_S^{(\infty)}$ and  $\delta c_S^{(k)}\equiv c_S^{(k)}-c_S^{(\infty)}$.

If the vector $\mathbf{X}\equiv\left(\delta p_S^{(k)},\delta c_S^{(k)},(\delta c_S^{(k)})^*\right)^T$ is considered, then Eq.~(\ref{eq:tracedistance_explicit}) describes a \textit{weighted 2-norm} with metrics given by $G=diag(1,1/2,1/2)$ \cite{VanLoan2013,Jiang2008,Sun2017}.

The time evolution of the vector $\mathbf{X}$ is governed by the map $\mathbf{X}^{(k+1)}=M\mathbf{X}^{(k)}$, with the matrix M given in Eq.~(\ref{eq:cohancillas_M-matrixevolution}). Thus, $\mathbf{X}^{(k)}=M^k\mathbf{X}^{(0)}$.
Considering the distance,
\begin{align}
D\left(\rho_S^{(\infty)},\rho_S^{(k)}\right)&=||M^k\mathbf{X}^{(0)}||\\
    &\leq ||M||^k||\mathbf{X}^{(0)}||
=||M||^k D\left(\rho_S^{(\infty)},\rho_S^{(0)}\right),
\end{align}
with the norm induced on the matrix $M$ given by
\begin{align}
    ||M||&\equiv\sqrt{\lambda_{max}(G^{-1}M^\dagger GM)}\nonumber\\
    &=\sqrt{1-\sin^2(2J\tau)\left(1-4|c_A|^2\right)},
    \label{eq:Mnorm}
\end{align}
where $\lambda_{max}(G^{-1}M^\dagger GM)$ indicates the greatest eigenvalue of the matrix $G^{-1}M^\dagger GM$ \cite{VanLoan2013}. 

For $0<\|M\|<1$, a sufficient condition for the trace distance to be
smaller than $\epsilon$ is therefore
\begin{equation}
\|M\|^{n^*}
D\left(\rho_S^{(0)},\rho_S^{(\infty)}\right)
<
\epsilon.
\end{equation}
Taking the logarithms of both members in the last inequality, and replacing the expression of $||M||$, Eq.~(\ref{eq:Mnorm}), we get the threshold,
\begin{align}
n^{\rm bound}=\frac{\ln\left({\epsilon}\right)-\ln\left[{D(\rho_S^{(\infty)},\rho_S^{(0)})}\right]}{\ln\left[\sqrt{1-\sin^2(2J\tau)(1-4|c_A|^2)}\right]}.
    \label{eq:lowerbound}
\end{align}
This value serves as an upper bound on the minimum number of collisions required to guarantee convergence within the prescribed tolerance $\epsilon$,
\bea
n^*\leq n^{\rm bound}.
\eea
In Fig. \ref{fig:CoherentAncilla_StepNumber}, we illustrate that environmental coherence influences the number of iterations required for driving the system into a target coherent steady state. We consider three sets of $J$ and $\omega$, selected from Figure~\ref{fig:Temp_map}, two corresponding to the cooling of the system below the bath temperature and one to the heating. In general, increasing the coherence of the ancilla, $|c_A|$, increases the number of iterations required for convergence, indicating that environmental coherence slows down the relaxation dynamics. 
For small $|c_A|$, we observe an approximately linear increase in $n^*$ with coherence, while at larger $|c_A|$ the dependence becomes nonlinear. 
The upper bound is relatively tight at small \(|c_A|\) but becomes increasingly loose as the coherence grows.
In particular, the bound diverges as $|c_A|\rightarrow 1/2$, where the induced norm, $||M||$ approaches unity. This divergence does not necessarily imply a divergence in the actual convergence time.

\begin{figure*}[ht!]
    \centering
    \includegraphics[width=1\linewidth]{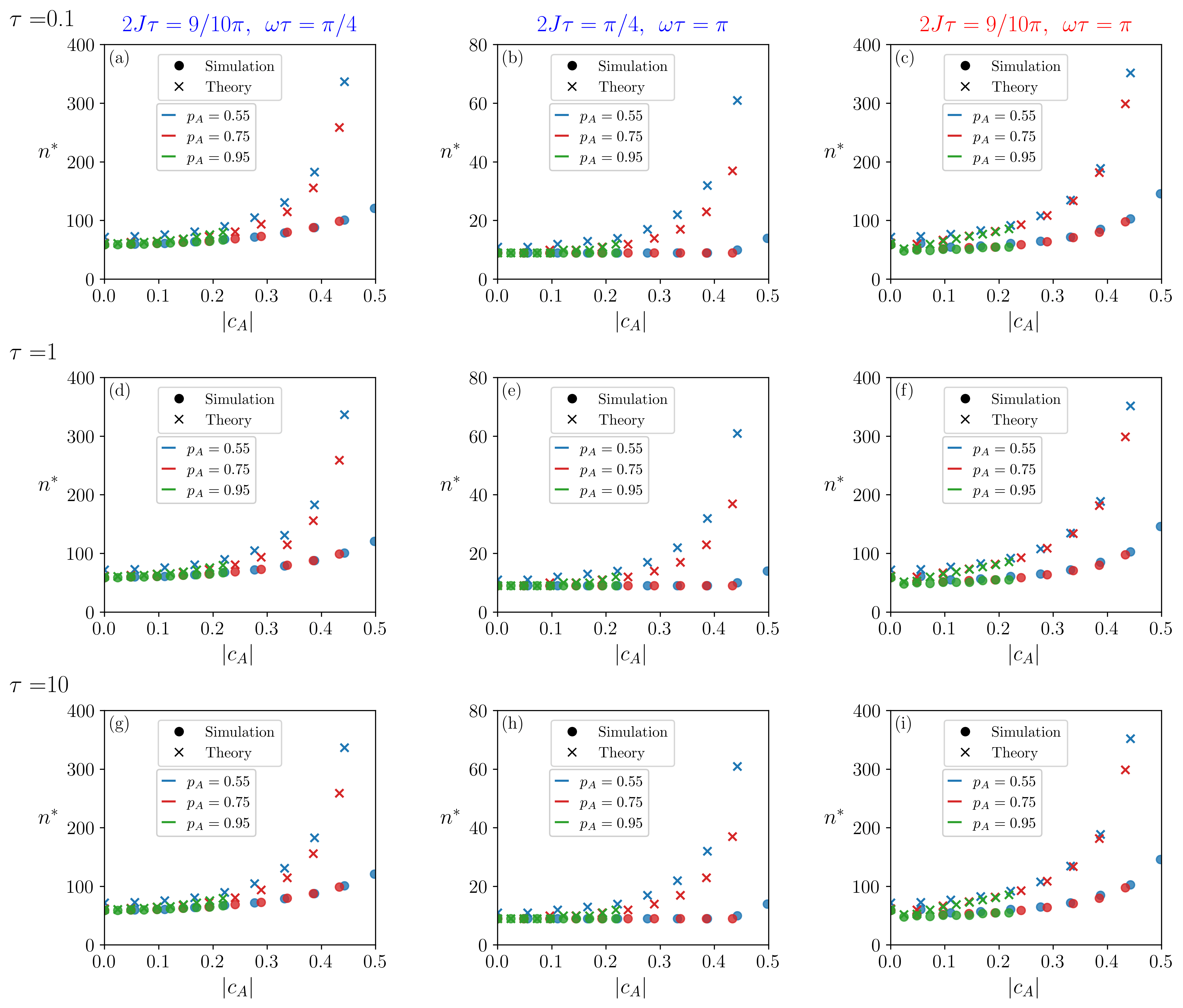}
    \caption{Number of steps required for preparing the system in a target  steady state with coherence as a function of the amount of environmental coherence. We chose three points from Fig. \ref{fig:Temp_map}, two for cooling (left  and center columns) and one for heating (right column), represented by specific values of $2J\tau$ and $\omega\tau$. The theoretical points in the plot are obtained using Eq.~(\ref{eq:lowerbound}). The initial state of the system is set by $p_S^{(0)} = 0.86$ and $c_S^{(0)} = -0.11-i0.15$. We considered three different durations for the single RI step, $\tau=0.1,1$, and $10$, and set $\epsilon=0.01$.}
    \label{fig:CoherentAncilla_StepNumber}
\end{figure*}

\section{Quantifying coherence}
\label{sec:QC}
In this Appendix, we apply the definitions of  the \textit{l1-norm} and the \textit{relative entropy of coherence} \cite{Plenio2014} to quantify the quantum coherence available as a resource that are stored in the steady state of the system, Eq. (\ref{eq:coherentancilla_exact_coh_ss}).
 
The \textit{l1-norm} of coherence is defined as
\begin{equation}
    C_{l_1}(\rho) = \sum_{\substack{i,j \\ i\neq j}} |\rho_{i,j}|.
    \label{eq:l1-metric}
\end{equation}

The \textit{relative entropy of coherence} is defined as
\begin{equation}
    C_{rel.ent.}(\rho) = S(\rho_{diag})-S(\rho),
    \label{eq:relent_metric}
\end{equation}
where $S(\rho)=-\text{Tr}(\rho\ln\rho)$ is the Shannon-Von Neumann entropy.

In Figures \ref{fig:l1norm} and \ref{fig:REC}, we report the l1-norm and the relative entropy of coherence, respectively, at the system steady state for varying degrees of environmental coherence. Overall, increasing the environmental coherence increases the amount of coherence available as a resource. However, this behavior is not always monotonic. In some cases, we observe a saturation while in others, the coherence stored in the system steady state available as a resource reaches an optimum at an intermediate level of environmental coherence and subsequently decreases as the environmental coherence is further increased. 

\begin{figure*}
    \centering
    \includegraphics[width=1\linewidth]{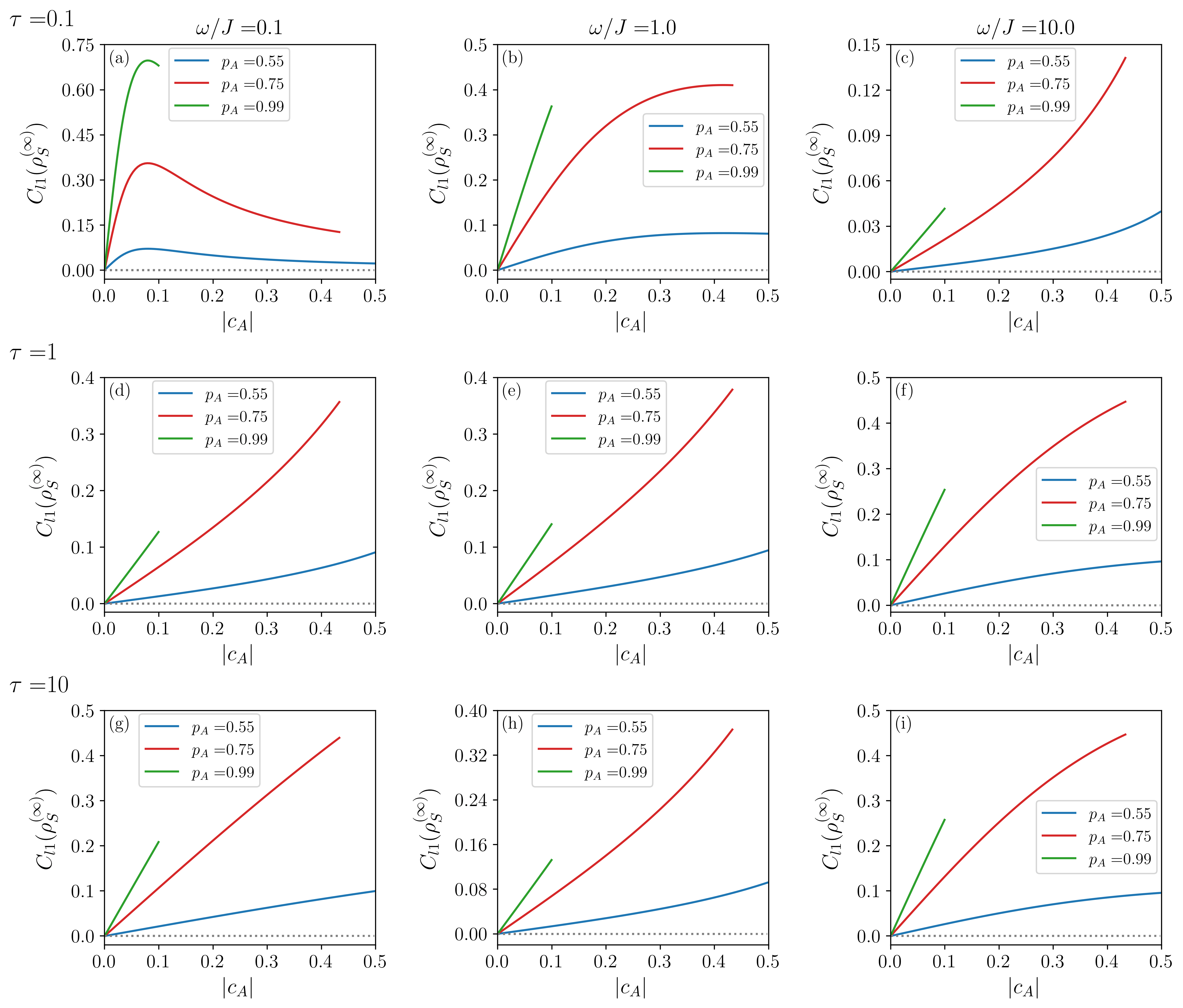}
    \caption{System coherence at steady state, quantified by the $l1-norm$, as a function of the environmental coherence for three different durations of the single collision, $\tau=0.1,1$, and $10$. We applied the definition in Eq.~(\ref{eq:l1-metric}) to the expression for the steady-state coherences, Eq.~(\ref{eq:coherentancilla_exact_coh_ss}).}
    \label{fig:l1norm}
\end{figure*}
\begin{figure*}
    \centering
    \includegraphics[width=1\linewidth]{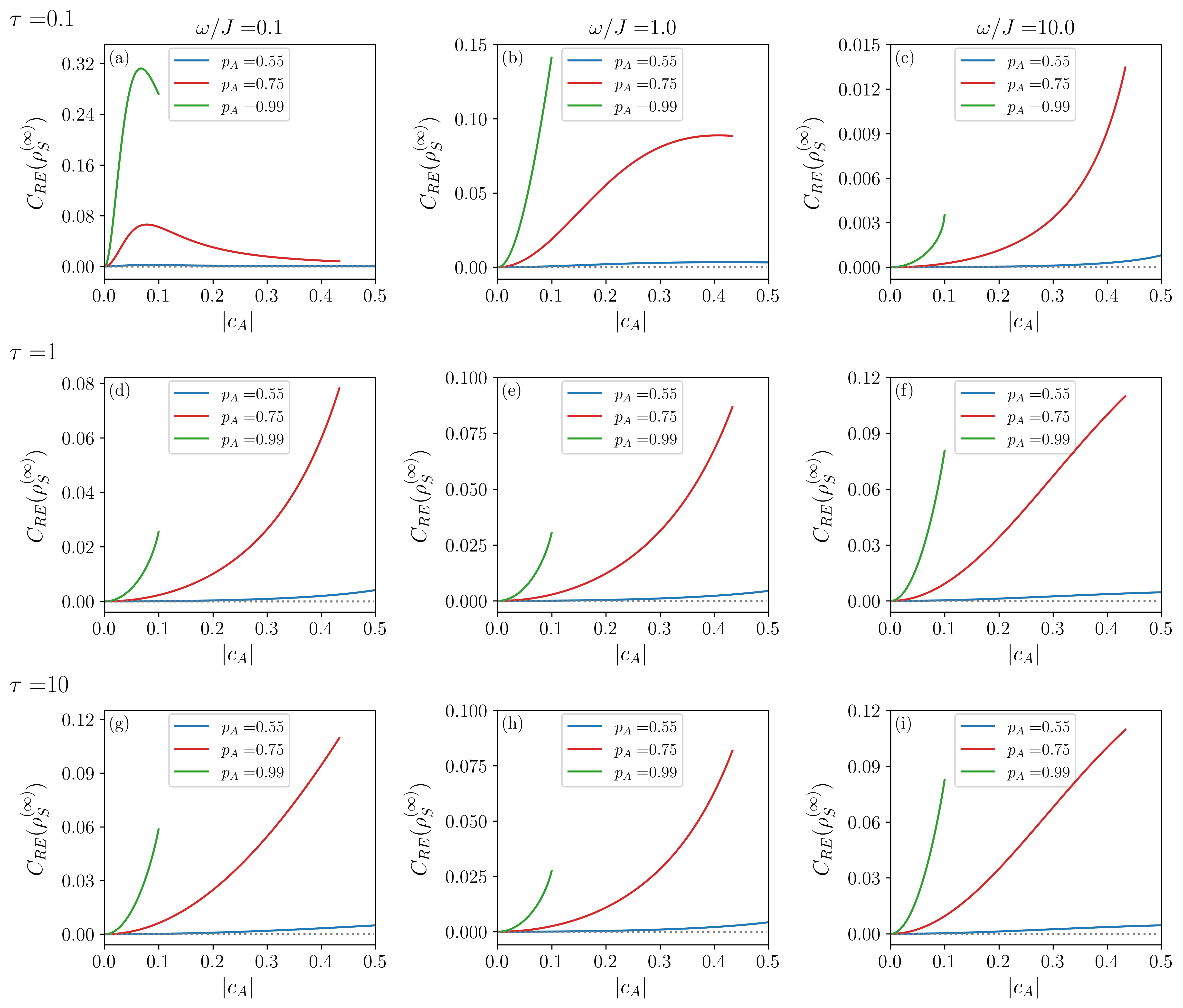}
    \caption{System coherence at steady state, quantified by the \textit{relative entropy of coherence}, as a function of the environmental coherence for three different durations of the single collision, $\tau=0.1,1$, and $10$. We applied the definition in Eq.~(\ref{eq:relent_metric}) to the expression for the steady-state coherence, Eq.~(\ref{eq:coherentancilla_exact_coh_ss}).}
    \label{fig:REC}
\end{figure*}

\end{widetext}

\end{document}